\documentclass[twocolumn]{aastex631}

\newcommand\imc{\texttt{imcascade}}
\newcommand\MLR{$M/L$}

\received{XXX}
\revised{YYY}
\accepted{ZZZZ}

\submitjournal{ApJ}

\shorttitle{Color gradients and half-mass radii out to $z=2$}
\shortauthors{Miller et al. }
\graphicspath{{./}{figures/}}

\begin{document}

\title{Color gradients and half-mass radii of galaxies out to $z=2$ in the CANDELS/3D-HST fields: further evidence for important differences in the evolution of mass-weighted and light-weighted sizes }

\correspondingauthor{Tim B. Miller}
\email{tim.miller@yale.edu}

\author[0000-0001-8367-6265]{Tim B. Miller}
\affiliation{Department of Astronomy, Yale University, 52 Hillhouse Ave., New Haven, CT, USA, 06511}

\author[0000-0002-8282-9888]{Pieter van Dokkum}
\affil{Department of Astronomy, Yale University, 52 Hillhouse Ave., New Haven, CT, USA, 06511}

\author[0000-0002-8530-9765]{Lamiya Mowla}
\affil{Dunlap Institute for Astronomy and Astrophysics, University of Toronto, 50 St George Street, Toronto ON, M5S 3H4, Canada }

\begin{abstract}
Recent studies have indicated that the ratio between half-mass and half-light radii, $r_{\rm mass} / r_{\rm light}$, varies significantly as a function of stellar mass and redshift, complicating the interpretation of the ubiquitous $r_{\rm light}- M_*$ relation. To investigate, in this study we construct the light and color profiles of $\sim 3000$ galaxies at $1<z<2$ with $\log\, M_*/M_\odot > 10.25$ using  \texttt{imcascade}, a Bayesian implementation of the Multi-Gaussian expansion (MGE) technique. \imc{} flexibly represents galaxy profiles using a series of Gaussians, free of any a-priori parameterization. We find that both star-forming and quiescent galaxies have on average negative color gradients. For star forming galaxies, we find steeper gradients that evolve with redshift and correlate with dust content. Using the color gradients as a proxy for gradients in the $M/L$ ratio we measure half mass radii for our sample of galaxies. There is significant scatter in individual $r_{\rm mass} / r_{\rm light}$ ratios, which is correlated with variation in the color gradients. We find that the median $r_{\rm mass} / r_{\rm light}$ ratio evolves from 0.75 at $z=2$ to 0.5 at $z=1$, consistent with previous results. We characterize the $r_{\rm mass}- M_*$ relation and we find that it has a shallower slope and shows less redshift evolution than the $r_{\rm light} - M_*$ relation. This applies both to star-forming and quiescent galaxies. We discuss some of the implications of using $r_{\rm mass}$ instead of $r_{\rm light}$, including an investigation of the size-inclination bias and a comparison to numerical simulations.

\end{abstract}


\section{Introduction} \label{sec:intro}

The physical processes that affect galaxies shape their morphology. A crucial tool to understand the build-up of stellar mass within galaxies is the relation between galaxy size and stellar mass. Many studies, spanning a wide range of observational facilities and galaxy properties, have produced a consistent picture of the evolution of galaxy structure from high-to-low redshift \citep{Shen2003,Ferguson2003, Trujillo2006, Williams2010, Ono2013,vanderWel2014,Lange2015,Mowla2019,Kawinwanichakij2021}. A galaxy's half-light radius, also known as the effective radius or $r_{\rm light}$, is most often used as a measurement of its size \citep[although some alternatives have been proposed; see, e.g.,][]{Ribeiro2016,Miller2019,Trujillo2020}. The slope of the $r_{\rm light}$ - $M_*$ relation is generally observed to be positive, i.e., more massive galaxies are larger, and star-forming galaxies are seen to have have larger sizes than their quiescent counterparts in almost all circumstances, except at the highest masses ($\log M*/M_\odot \gtrsim 11$ \citep{Shen2003,Mowla2019}). 

The slope of the size-mass relation for star-forming galaxies is measured to be fairly shallow, $d\,\log\,r_{\rm light} / d\,\log\,M_* \sim 0.2$, suggesting that growth is largely self-similar, where star formation proceeds equally at all radii. For quiescent galaxies, a much steeper slope is observed along with rapid evolution in size since $z=2$. These observations have been linked to growth via minor mergers; by depositing remnants of galaxies at large radii and disturbing the central galaxy, minor mergers increase the radius of a galaxy proportionally more than the stellar mass \citep{Naab2009,Bezanson2009,Newman2012,VanDokkum2015}. A complication is  progenitor bias, as galaxies that cease their star formation later introduce a bias into the observed evolution of the size evolution of quiescent galaxies \citep{vanDokkum2001,Carollo2013,Poggianti2013}.

An important caveat to all of these studies is that they are not based on half-{\em mass} radii but on half-{\em light} radii, which are typically measured at rest-frame optical wavelengths. This produces a biased view of the underlying stellar mass distribution as galaxies are known to have color gradients. Gradients in the color are the manifestation of differing stellar populations or dust content which affect the mass-to-light ratios (\MLR{}). Spatial variations in \MLR{} can bias light weighted morphology away from the intrinsic stellar mass distribution within galaxies. To study the stellar mass build up of galaxies, the mass-weighted radius is a more fundamental measurement.

\MLR{} gradients are traced by color gradients which have been observed in both star-forming and quiescent galaxies in the local universe since the adoption of the CCD camera in astronomical observations~\citep{Kormendy1989,Franx1990,Balcells1994}. For both types of galaxies color gradients are on average negative, i.e., going from a redder center to bluer outskirts. For star-forming galaxies this is typically due to the presence of distinct components, namely an old red bulge in the center and a young blue disk at larger radii. For quiescent galaxies the main cause is likely a lower metallicity in the outskirts ~\citep{Wu2005,LaBarbera2009,Tortora2011}, with perhaps some contribution from variation in age and/or stellar initial mass function \citep{Conroy2017,DominguezSanchez2019,VanDokkum2021}.

Whatever the cause of the color gradients, negative color gradients imply negative \MLR{} gradients, as variations in age, metallicity, and dust content all produce approximately the same variation in the plane of \MLR{} ratio and color \citep{Bell2001}.
As a result, consistently negative color gradients imply that on average galaxies' half-mass radii, $r_{\rm mass}$, are systematically smaller than their optical half-light radii. In the local universe the median $r_{\rm light}$ at near-IR wavelengths, a more direct tracer of stellar mass than optical, is indeed observed to be smaller than at optical wavelengths~\citep{Mollenhoff2006,Kelvin2012,Lange2015}. Similarly, studies directly investigating $r_{\rm mass}$ have shown it to be smaller than $r_{\rm light}$ at $z\sim 0$ \citep{LaBarbera2009,Ibarra-Medel2022}.

Initial studies  at $z\gtrsim 1$ have mirrored these results, at least qualitatively. At high redshift average color gradients have been observed to be negative for all types of galaxies \citep{Guo2011,Wuyts2013,Szomoru2013,Liu2016,Wang2017,Liu2017,Suess2019}. Furthermore, \citet{Szomoru2013} studied 177 massive galaxies ($\log M_*/ M_\odot > 10.7$) and found $r_{\rm mass}$ was on average $25\%$ smaller than $r_{\rm light}$, which did not correlate significantly with redshift or any galaxy properties. \citet{Chan2016} find $r_{\rm mass}$ is $41 \%$ smaller than $r_{\rm light}$ for 36 quiescent galaxies in a $z=1.55$ cluster. A crucial aspect of these studies is the finding that the ratio of $r_{\rm mass}/r_{\rm light}$ did not depend significantly on galaxy properties, such as stellar mass or $r_{\rm light}$. In this scenario, the interpretations of the $r_{\rm light}-M_*$ relation would remain intact as converting to $r_{\rm mass}$ would only entail a constant scaling from the optical half-light radii. 

This situation changed with the results of \citet{Suess2019}, who measured half-mass radii for a large, uniform sample of $\sim 7000$ galaxies in the CANDELS fields with a wide mass range of $\log M_* = 9.5-11.5$ at $z=1-2.5$. With this sample, they find significant correlations between $r_{\rm mass}/r_{\rm light}$ and many galaxy properties for both star-forming and quiescent galaxies. Importantly, they find that the median ratio of $r_{\rm mass}/r_{\rm light}$ evolves significant between $z=1-2.5$, such that the median half-mass radii of galaxies do not evolve strongly between $z=1-2.5$. \citet{Mosleh2020} analyzed $r_{\rm mass}-M_*$ distribution for galaxies between $0.3<z<2$ using an independent method to measure half-mass radii. They also find little evolution in the $r_{\rm mass}-M_*$ relation for either star-forming or quiescent galaxies across the redshift range. These recent results show that \MLR{} gradients may mimic evolution in the $r_{\rm mass}-M_*$ and alter its appearance. The observed lack of evolution in the median $r_{\rm mass}$ of galaxies is starkly different from the evolution of $r_{\rm light}$.

There is some tension between the earlier results of \citet{Szomoru2013} and \citet{Chan2016} and the later results of \citet{Suess2019} and \citet{Mosleh2020}, and to understand the structural evolution of galaxies it is imperative to come to a firm measurement of the evolution of \MLR{} gradients and how they alter our interpretation of $r_{\rm light}$. The tension in the literature may be caused by differences in sample size and selection \citep[see][for a discussion]{Suess2019}, by differences in methodologies or how they were applied, or a combination of effects. Specifically, all these studies rely on parametric fits to model the light profile and/or the \MLR{} profile, and this can induce systematic errors if reality does not match the parametric form. 

As an alternative to parametric methods we recently developed \imc{}, a Bayesian implementation of the Multi-Gaussian expansion (MGE) formalism~\citep{Miller2021}. MGE flexibly models galaxies as a mixture of Gaussians and therefore does not require the a-priori choice of parameterization. In \citet{Miller2021} we show that \imc~can accurately model faint and semi-resolved galaxies, like those at $z\sim1.5$ in HST images. 

In this paper we use \imc~applied to data from the Cosmic Assembly Near-infrared Deep Extragalactic Legacy Survey (CANDELS) and 3D-HST surveys \citep{Koekemoer2011,Skelton2014} to measure color gradients and half-mass radii for galaxies at $1<z<2$. By using \imc~to model the galaxy light distribution and account for the PSF, we measure intrinsic light and color profiles free of any parameterization. We then measure half-mass radii by using the color profile to account for variations in \MLR{}. In this study we focus on the evolution of color gradients along with how they affect the relationship between $r_{\rm light}$ and $r_{\rm mass}$.

The paper is organized as follows. In Section~\ref{sec:data} we describe our galaxy sample, the 3D-HST and CANDELS data that are used, and how we model the images with \imc. Section~\ref{sec:col_grad} contains results showing the evolution of observed color gradients, Section~\ref{sec:hl_hm} discusses the relationship between $r_{\rm light}$ and $r_{\rm mass}$ and in Section~\ref{sec:size-mass} we analyze the resulting $r_{\rm mass}-M_*$ relation, focusing on how it compares to the well-studied $r_{\rm light}$ version. Our method and the results are discussed in Section~\ref{sec:disc} along with our final conclusions in~\ref{sec:conc}.

Throughout this paper we assume a cosmology with $H_0 = 68\ {\rm km/s/Mpc}$ and $\Omega_m = 0.31$ following \citet{Planck2015}. Unless otherwise stated, ``radius''  in this paper represents the semi-major half-light (or half-mass) radius. Surface brightness, color, and surface density profiles are shown along the semi-major axis.

\section{Observations and Modeling} \label{sec:data}
\begin{figure*}
    \centering
    \includegraphics[width = \textwidth]{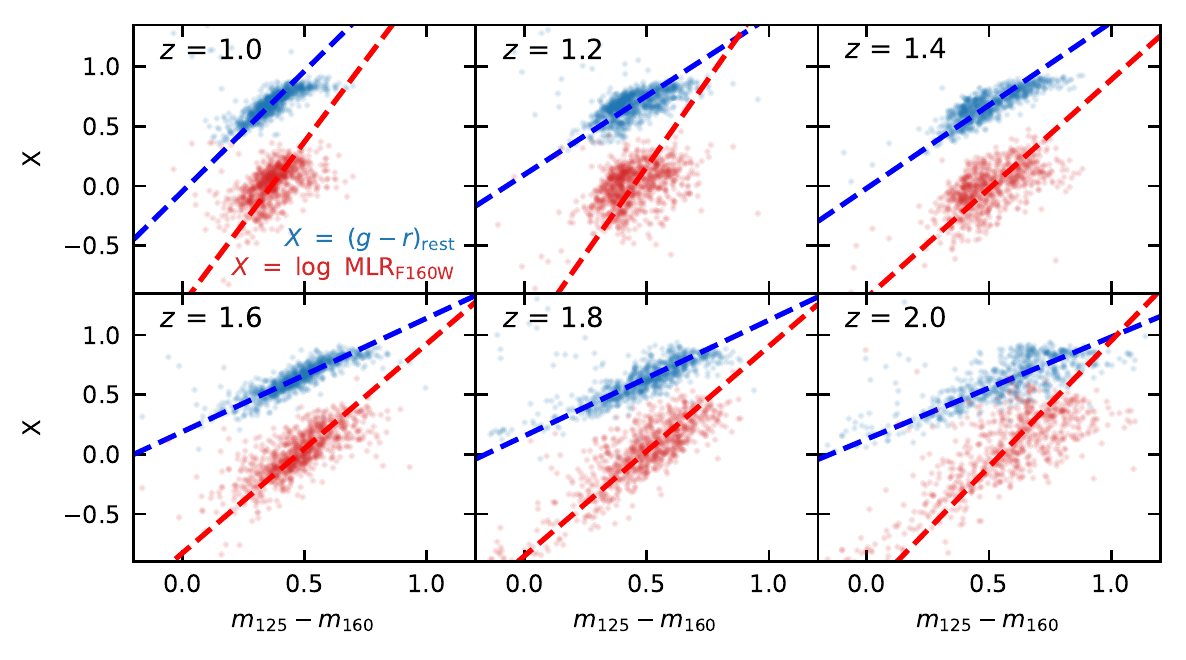}
    \caption{Linear fits that are used in this paper to convert the observed color to rest frame $g-r$ color and mass-to-light ratio. Fits are performed to results from EAZY and FAST as derived in \citet{Skelton2014} for the integrated light of galaxies in the 3D-HST survey.}
    \label{fig:col_fit}
\end{figure*}

\subsection{Galaxy Sample and Observations}
We derive the galaxy sample for this study from the 3D-HST catalog~\citep{Brammer2012, Skelton2014}. This catalog uses \texttt{FAST} to derive stellar masses and \texttt{EAZY} to derive rest frame colors ~\citep{Kriek2018,Brammer2008}. We select all galaxies in the catalog with $\log\ M_*\ >\ 10.25$, $1 < z < 2$, SNR$_{\rm F125W} > 25$ and SNR$_{\rm F160W} > 25$, which is equivalent to $m_{F160W} < 23$ in the tests performed in \citet{Miller2021} This sample is $>90\%$ mass complete. The signal-to-noise ratio for each band is measured within a 0.7 arcsec aperture. This sample contains a total of 3481 galaxies. As a supplement we use the size catalog from \citet{vanderWel2012}, who measured structural parameters of the galaxies using Sersic models. Galaxies are separated into star-forming and quiescent based on their rest frame $U,V$ and $J$ colors following the prescription in ~\citet{Muzzin2013}.

The relatively high mass cutoff of our sample is due to \imc~ requiring a higher SNR to accurately model galaxies than parameterized fitting techniques such as Sersic fitting with \texttt{galfit}, \texttt{imfit}, etc. This is a downside of the more flexible MGE method implemented in \imc. In \citet{Miller2021} we perform injection-recovery tests and find reliable results down to $m_{F160W} \sim 23$ for the depth of the CANDELS/3D-HST survey. This is contrasted with fitting single Sersic profiles to galaxies which has been shown to be reliable down to $m_{F160W} \sim 24$ in CANDELS data ~\citep{vanderWel2012}.

In this study we utilize HST data from the five fields (UDS, AEGIS, COSMOS, GOODSS and GOODSN) studied in the CANDELS survey~\citep{Koekemoer2011,Grogin2011}. We use the mosaics produced by the 3D-HST survey\footnote{The mosaics were downloaded from the 3D-HST website: https://archive.stsci.edu/prepds/3d-hst/} \citep{Skelton2014}. We utilize the F160W and the PSF-matched F125W mosaic. We opt for the PSF-matched images in order to help reduce artificial wiggles in the brightness and color profiles that can occur when using a discrete number of Gaussians in the MGE techniques \citep[see Fig.\ 2 in][]{Miller2021}.

While HST imaging exists at shorter wavelengths, specifically F814W and F606W for most of these fields, we opt to focus on F125W and F160W for this study. This ensures uniformity of analysis for all galaxies in our sample. Essentially all ($>95\%$) galaxies that meet our SNR criterion in F160W are also bright enough in F125W. For comparison, only one third of galaxies in our sample with SNR$_{F160W} > 25$ meet that criterion in F814W. This is due to a combination of intrinsic variation of rest-frame UV luminosity for galaxies at this epoch and the slightly different coverage of the optical and near-IR filters.

\begin{figure}
    \centering
    \includegraphics[width = \columnwidth]{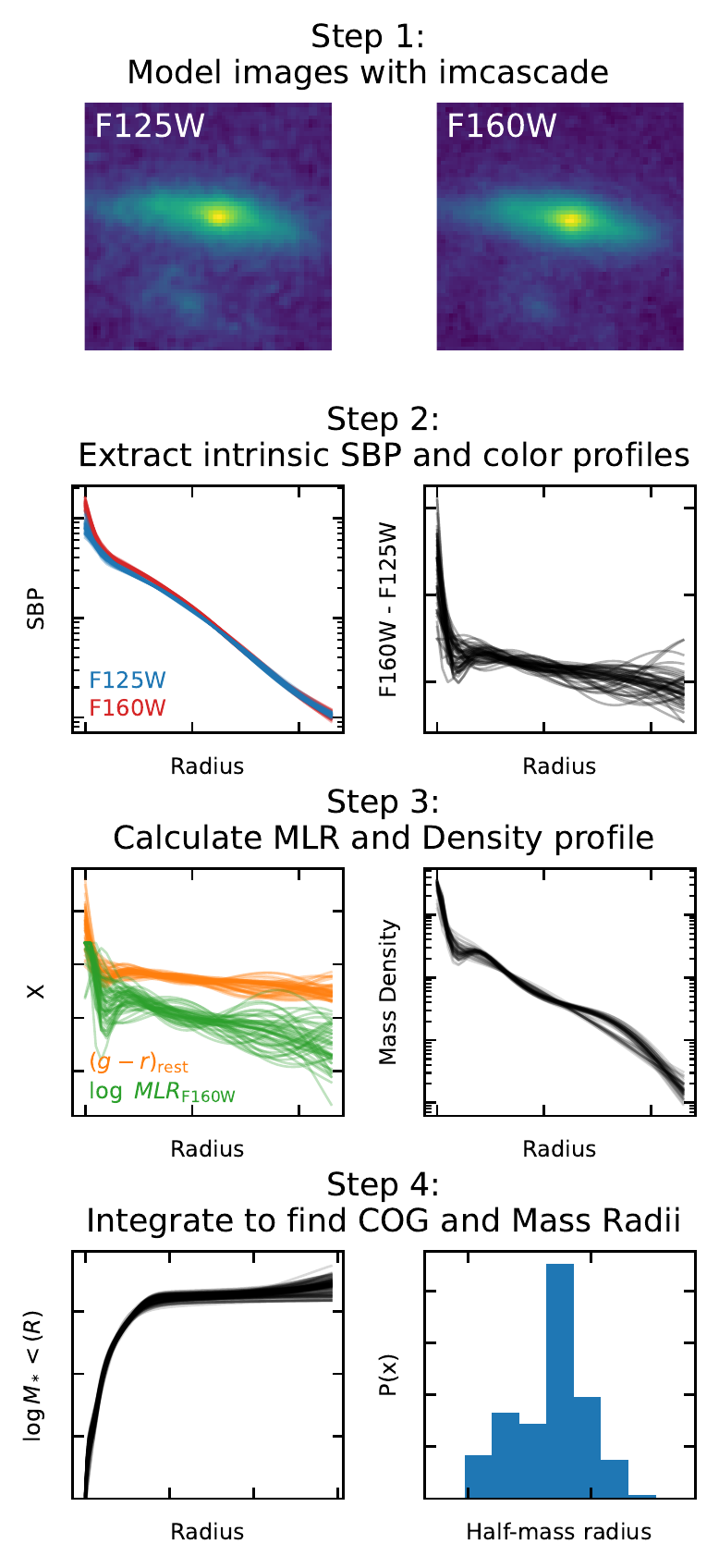}
    \caption{An overview of our methods, described in Section ~\ref{sec:data}, that we use to measure a galaxy's color profile and half-mass radius. }
    \label{fig:meth}
\end{figure}

\subsection{\imc~Modeling}

To model the light distibution of galaxies we use our recently developed method~\imc\ \citep{Miller2021}. \imc~is a Bayesian implementation of the Multi-Gaussian Expansion (MGE) method \citep{Emsellem1994a,Emsellem1994b, Cappellari2002}. MGE models galaxies as a mixture of Gaussians, allowing for increased flexibility over traditional parameterized techniques. Additionally, using a Gaussian decomposition of the PSF, this allows for analytic convolution, increasing performance and accuracy. In ~\imc, the widths (or standard deviations) of all the Gaussian components are fixed and the fluxes (or weights) are the free parameters that are fit for. Additionally the central position, position angle and axis ratio are fit for, but set to be the same for all components. 

Bayesian inference with~\imc\ is performed in two steps. The first step is a least squares fit to the image, varying all the parameters: the fluxes of each component along with the central position, position angle and axis ratio. After this the structural parameters, central position, position angle, and axis ratio are fixed. In \citet{Miller2021} we show that the least squares fitting process consistently and accurately returns these parameters. With these fixed, images of each component can be pre-rendered. Then model generation is a simple sum of existing array which is orders of magnitude faster computationally compared to rendering each component individually. With this performance increase Bayesian inference can be reasonably performed on a population of galaxies. Inference is performed with nested sampling using \texttt{dynesty}\ \citep{dynesty}, to derive the posteriors of the weights along with the Bayesian evidence. 

Another technique we employ is Bayesian model averaging described in \citet{Miller2021}. Specifically since the choice of widths for the Gaussian components is not unique, we perform inference on the same galaxy using multiple sets of different widths. The derived posteriors are then combined according to the resulting Bayesian evidences. We show that this helps reduce artificial ``wiggles" in the derived profiles resulting from the discrete number of Gaussian components used. 

For a full description of the \imc~method and its implementation please see \citet{Miller2021} in which we introduce and fully describe it. The procedure we use here follows that described in Sec.\ 4.4 of \citet{Miller2021}. That paper also includes injection-recovery tests which show that we can accurately model realistic $z\sim1.5$ galaxies in CANDELS-like imaging. 

To model our galaxies, we begin by preparing cutouts of each galaxy for both F160W and F125W with size $40\times r_{\rm eff, F160W}$, taken from the \citet{vanderWel2012} size catalog. We then create a variance image, taking into account the instrumental response and Poissson noise. A mask for each galaxy is created using the segmentation map provided by the 3D-HST team, and derived using a combination of the F125W, F140W and F160W image.

Each galaxy image is modelled with \imc~using five sets of ten Gaussian widths. This initial set of Gaussian widths is logarithmically spaced from $0.5$ pixel to $10\ \times\ r_{\rm vdW2012}$, utilizing the F160W size measured by \citet{vanderWel2012}. The subsequent sets are then shifted in logarithmic space by integer factors of $d\log \sigma / 5$, one fifth the logarithmic spacing between widths in the initial set. This creates a set of widths which are uniformly spaced in $\log\, r$. The same sets of widths for the Gaussian components are used for each galaxy to fit both the F160W and F125W images. 

The posterior distributions for the weights (or fluxes) of each component and the Bayesian evidence are calculated through nested sampling using \texttt{dynesty} \citep{dynesty}, as described above and in \citet{Miller2021}. For each image of each galaxy we then combine the posteriors of the observed quantities (flux, half-light radii and surface brightness profile) derived for each set of widths weighted by the relative evidences following the Bayesian model averaging method. These combined posteriors are then used in the further analysis.

\begin{table*}[]
    \centering
    \caption{Measurements of half-light radii, half-mass radii and color gradients using \imc{}}
    \begin{tabular}{cccccccccc}
 Field & ID$^*$ & RA$^*$ & DEC$^*$ & z$^*$ & $\log M_{\rm star}/M_\odot ^\dagger$ & Quenched$^\bot$ & $r_{\rm light}$ (kpc)$^\ddag$ & $r_{\rm mass}$ (kpc) $^\ddag$ & $d\, (g-r)\, /\, d\, \log r^\ddag$  \\ \hline \hline
GOODSN & 12078 & 189.167 & 62.202 & 1.017 & 10.560 & False & $1.788 \pm 0.014$  & $2.430 \pm 0.029$  & $0.361 \pm 0.115$  \\
GOODSN & 9766 & 189.172 & 62.192 & 1.096 & 10.710 & False & $6.914 \pm 0.137$  & $3.884 \pm 0.060$  & $-0.461 \pm 0.066$  \\
GOODSN & 2684 & 189.092 & 62.147 & 1.016 & 10.670 & True & $1.803 \pm 0.026$  & $1.054 \pm 0.080$  & $-0.457 \pm 0.338$  \\
GOODSN & 36408 & 189.361 & 62.338 & 1.006 & 10.350 & True & $0.760 \pm 0.029$  & $0.452 \pm 0.089$  & $-0.543 \pm 0.387$  \\
GOODSN & 32841 & 189.099 & 62.310 & 1.004 & 11.020 & True & $11.079 \pm 0.490$  & $1.972 \pm 0.075$  & $-1.236 \pm 0.280$  \\
GOODSN & 34999 & 189.249 & 62.326 & 1.146 & 10.520 & False & $2.828 \pm 0.061$  & $1.446 \pm 0.132$  & $-0.207 \pm 0.143$  \\
GOODSN & 9861 & 189.317 & 62.192 & 1.081 & 10.780 & True & $4.510 \pm 0.088$  & $1.322 \pm 0.340$  & $-0.194 \pm 0.124$  \\
GOODSN & 5386 & 189.140 & 62.168 & 1.016 & 10.770 & False & $6.274 \pm 0.031$  & $1.903 \pm 0.030$  & $-0.867 \pm 0.086$  \\
GOODSN & 8560 & 189.389 & 62.184 & 1.115 & 10.770 & True & $2.111 \pm 0.127$  & $1.305 \pm 0.117$  & $-1.350 \pm 0.435$  \\
GOODSN & 5584 & 189.063 & 62.169 & 1.027 & 10.980 & False & $3.702 \pm 0.081$  & $2.521 \pm 0.176$  & $-0.117 \pm 0.129$  \\
... & ... & ... & ... & ... & ... & ... & ... & ... & ... \\
\hline
\end{tabular}
{\footnotesize
\raggedright
$^*$ Taken from 3D-HST catalog \citep{Skelton2014}\\
$^\dagger$ Taken from 3D-HST catalog, calculated using FAST \citep{Skelton2014}\\
$^\bot$ According to rest-frame UVJ colors following \citet{Muzzin2013} \\
$^\ddag$ Median value from posterior. The uncertainty is half of the 16th-84th percentile range.\\
The radii are half-(light or mass) radii measured along the semi-major axis. \\ 
The color gradient is measured between $0.5-2\ r_{\rm light}$, see Sec.~\ref{sec:col_grad} for more details.\\
}
\label{tab:data_catalog}
\end{table*}

\begin{figure*}
    \centering
    
    \includegraphics[width =0.99\textwidth]{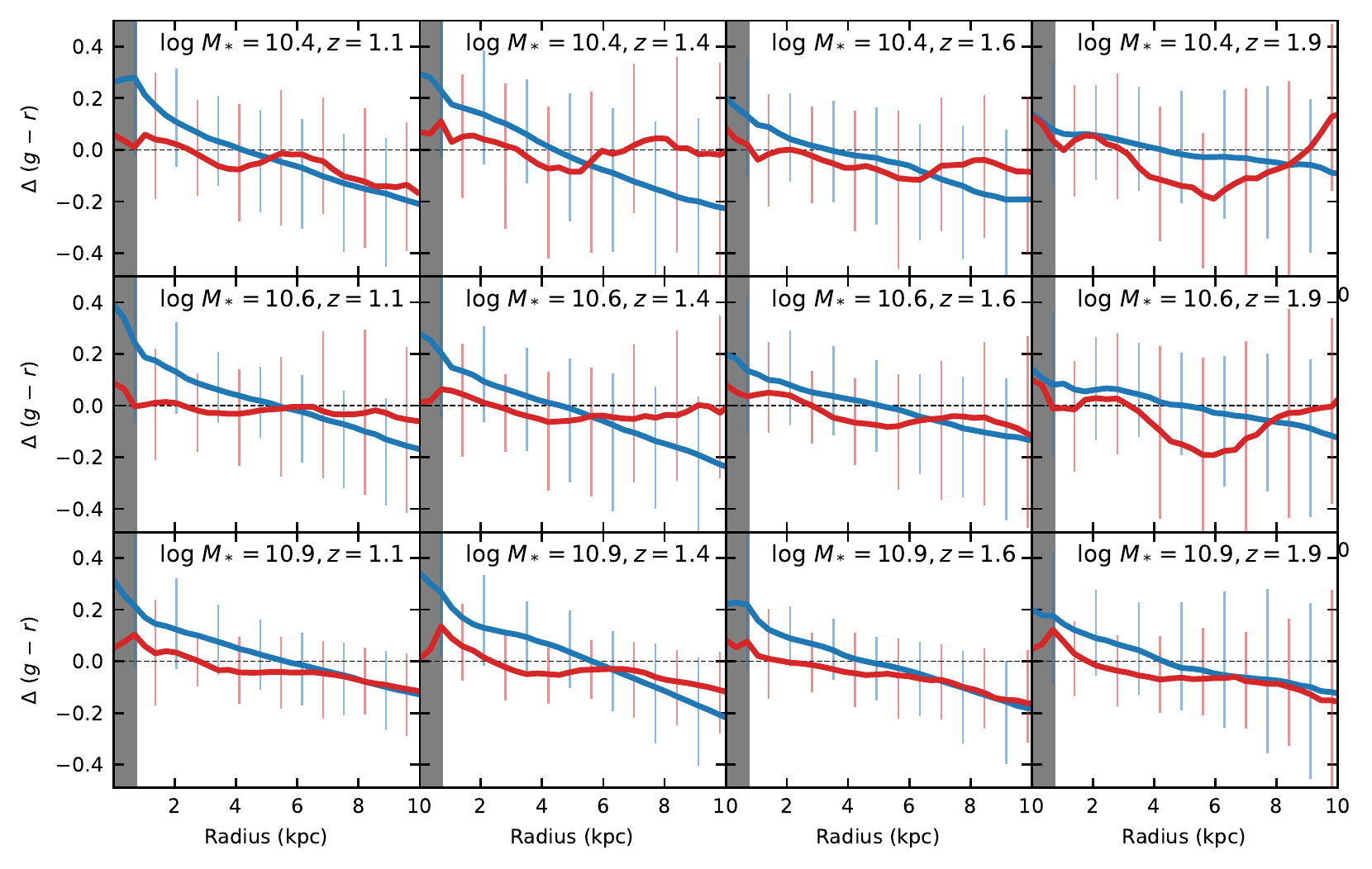}

    \caption{Median color profiles over a range of stellar masses and redshifts. Color profiles are normalized by each galaxy's total color before calculating the median. In each panel the blue line shows the median profile for star-forming galaxies and the red line for quiescent galaxies. Errorbars show the 16th - 84th percentile range, which encapsulates both observational uncertainties and intrinsic scatter with each population. Stellar mass increases top to bottom and redshift increases left to right with the mid point of each stellar mass and redshift bin also displayed in each panel. The thin dotted line shows where $\Delta (g-r) = 0$ and the grey region signifies where $r < $ PSF HWHM. In general star-forming galaxies have larger color gradients which get steeper at lower redshift and higher masses. We find very mild color gradients in quiescent galaxies that do not appear to correlate strongly with stellar mass or redshift.}
    \label{fig:all_col_prof}
\end{figure*}

\begin{figure}
    \centering
    \includegraphics[width = 0.95\columnwidth]{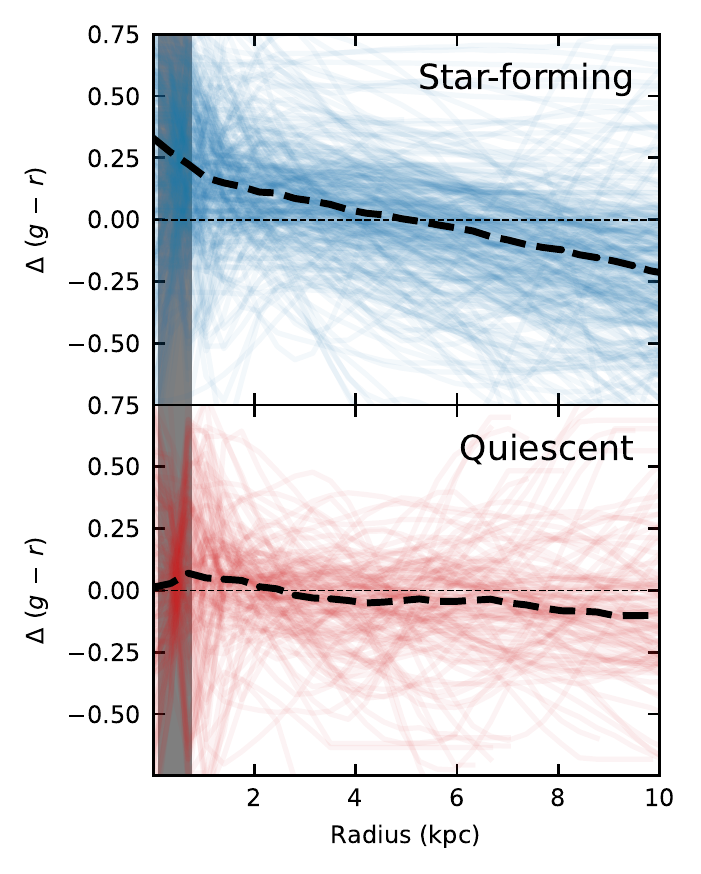}
    \caption{Normalized color profiles for individual galaxies with $\log M_*/M_\odot = 10.25-10.75$ at $z=1-1.25$. The median normalized color from the posterior distribution at each radius is used to construct the color profile. The top panel shows star-forming galaxies in blue and the bottom panel shows quiescent galaxies in red. In each panel the black dashed line shown the median of the galaxy sample. We see a large amount of scatter in the color profiles of both types of galaxies which tends to increase at smaller radii.}
    \label{fig:ind_col_prof}
\end{figure}

\subsection{Measuring mass-weighted radii}
\label{sec:hm_meas}
In order to calculate the mass-weighted radii and other parameters, we must convert the observed color profiles to physical quantities. We achieve this by deriving a relationship between the observed color, $m_{\rm F125W} - m_{\rm F160W}$ and $(M/L)_{\rm F160W}$ to convert brightness profiles to mass profiles. Additionally we convert the observed color to rest-frame $(g-r)$ so we can effectively compare color gradients across redshifts. These methods are necessarily approximate: as we show below, there is considerable scatter in $M/L$ ratio at fixed $m_{\rm F125W} - m_{\rm F160W}$ color. As explained in the Introduction, this paper is concerned with broad trends, and for this purpose our methodology is sufficient.\footnote{We can expect great improvements in this area from JWST, owing to its ability to obtain high resolution images at rest-frame near-IR wavelengths.}

In Figure~\ref{fig:col_fit} we show linear fits to $\log (M/L)_{\rm F160W}$ and $(g-r)_{\rm rest}$ as a function of observed $m_{\rm F125W} - m_{\rm F160W}$ color for galaxies in six redshift bins. These measurements are taken from the 3D-HST catalogs ~\cite{Skelton2014}. The best-fitting relations are shown in this figure and also in Appendix \ref{sec:obs_col_rel}. We place a floor and ceiling of $\log (M/L)_{\rm F160W} = -1$ and $0.7$ respectively. This will limit the effect of uncertain or outlying colors of altering the resultant half-mass radii. We will use these relationships to convert the observed color profiles measured by \imc~ to rest frame color and \MLR{} profiles.

\begin{figure*}
    \centering
    \includegraphics[width = \textwidth]{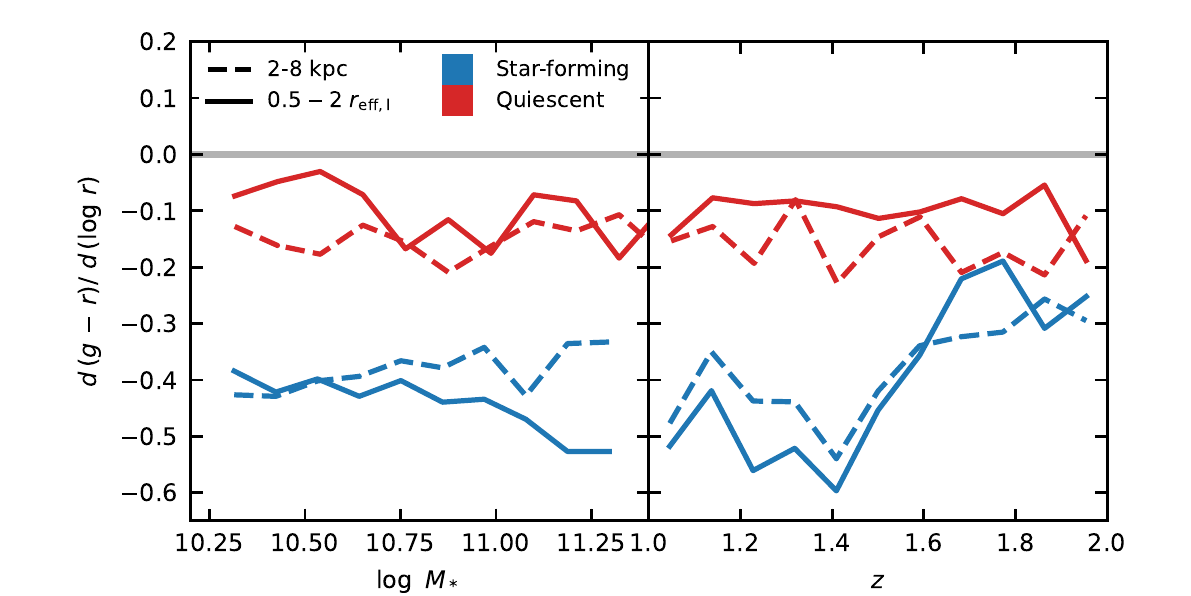}
    \caption{Measured color gradients shown as a function of redshift and stellar mass. Median color gradients for star-forming galaxies are shown in blue with quiescent galaxies in red. Additionally the gradient is measured in two separate ways: the solid line shows the gradient between $0.5-2 r_{\rm eff, l}$ and the dashed lines show the gradient measured between a fixed physical radius of $2-8$ kpc. Quiescent galaxies have weaker gradients ($d\, (g-r) / d \log\, r \sim 0.15$) which do not depend on stellar mass or redshift while star-forming galaxies have stronger gradients ($d\, (g-r) / d \log\, r \sim 0.4$) which get steeper at lower redshift.}
    \label{fig:dc_dr}
\end{figure*}

To measure $r_{\rm mass}$ we begin by multiplying the \MLR{} profile by the F160W surface brightness profile, measured using \imc, to calculate the mass density profile. We numerically integrate the mass density profile to find the enclosed mass profile and calculate the half-mass radius, within $10\times r_{\rm light}$. The resulting measurements are not sensitive to the precise choice for this cutoff. This process is repeated, using samples from the posterior distribution of the surface brightness profiles and color profiles to build the posterior distribution of half-mass radii. The reported half-mass radii and associated errors are calculated as the median and half the 16th-84th percentile range respectively. An overview of the procedure for an example galaxy is shown in Figure~\ref{fig:meth}.

Our method to measure half-mass radii resembles that presented in \citet{Szomoru2013} and Method 3 presented in \citet{Suess2019}. Their method relies on the residual-corrected Sersic profile\footnote{The method of residual correcting the best fit Sersic profile, developed in \citet{Szomoru2010} is meant to alleviate biases caused by the choice of parameterization.} in several bands and \texttt{EAZY} to construct a $(u-g)_{\rm rest}$ profile which is then converted to a \MLR{} profile using a linear relation. Other studies \citep{Wuyts2013,Suess2019, Mosleh2020} have instead opted to perform SED fitting on several bands in either pixels or annuli to improve the accuracy of the \MLR{} ratio.

The difference between our method and previous ones is that we do not assume a parametric form for the luminosity profile. In previous studies color profiles and $M/L$ profiles were derived using parameterized functions. In our method the color profile is derived directly, and can be studied meaningfully along with $r_{\rm mass}$. In appendix~\ref{sec:meth_comp} we show a detailed comparison between our measurements and those presented in \citet{Suess2019} and \citet{Mosleh2020}. In general, our measurements appear largely consistent with both studies. This comparison is discussed further in Section~\ref{sec:disc} and Appendix~\ref{sec:meth_comp}.

\begin{figure}
    \centering
    \includegraphics[width = \columnwidth]{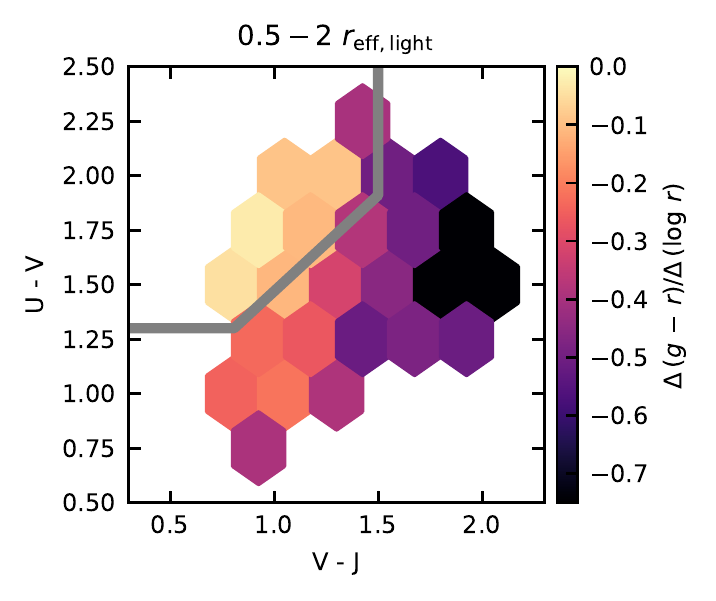}
    \caption{Color gradients across the $U-V$ vs.\ $V-J$  plane. The median color gradient within each hexbin is shown. Here we use the gradient measured between $0.5-2\ r_{\rm eff, l}$ but the results are similar if we use the gradient based on physical radii. For star-forming galaxies we see that the median color gradient increases from bottom-left to top-right, parallel to the track which is thought to trace increasing dust attenuation.}
    \label{fig:dcdr_UVJ}
\end{figure}

We make our measurements available online here\footnote{Found at this link: \url{https://raw.githubusercontent.com/tbmiller-astro/tbmiller-astro.github.io/main/assets/Miller2022_morph_CANDELs.txt} }. We include salient information taken from the 3D-HST catalog along with $r_{\rm light}$, $r_{\rm mass}$ and color gradient measurements as described in this section. For these measurement we include the mean of the posterior along with the uncertainty measured as half of the 16th-84th percentile range. We inspected random examples and found the posteriors to be roughly symmetric, making this a reasonable estimate of the width of the posterior. Table~\ref{tab:data_catalog} displays the first few entries of the table to show its form and content.

\section{Color Gradients}
\label{sec:col_grad}

Figure~\ref{fig:all_col_prof} displays median color profiles measured by \imc~for galaxies in our sample as a function of stellar mass and redshift. Here, and throughout we separate galaxies into star-forming and quiescent based on their rest frame UVJ color. Before calculating the median we normalize each galaxy's color profile by its integrated color and plot the normalized color profile: $\Delta (g-r)(R) = (g-r)(R) - (g-r)_{\rm int.}$. This allows us to focus on color gradients, changes in color as a function of radius, and enables easier comparison between galaxies which have a large range of integrated colors. Each panel shows the median $\Delta (g-r)$ for both star-forming and quiescent galaxies as a function of radius.

We observe several interesting trends in the median color profiles of galaxies. In general galaxies of all types tend to have negative color gradients: redder in the center and bluer in the outskirts. This agrees with the well established results in both the local universe and at high redshift ~\citep{Kormendy1989, Wu2005,Wuyts2013,Suess2019}. Star-forming galaxies appear to have stronger color gradients, i.e. larger negative slopes, compared to quiescent galaxies. Additionally for star-forming galaxies the slope of the color profiles is steeper at higher mass and lower redshift. This is especially evident for the central region, within $\sim 2\ $ kpc. Quiescent galaxies appear to have consistently negative but relatively shallower color profiles with no apparent trends with redshift and stellar mass.

Individual color profiles for galaxies at a fixed stellar mass and  redshift ($\log M_*/M_\odot = 10.25-10.75$ and $z=1-1.25)$ are shown in Figure~\ref{fig:ind_col_prof}. For each galaxy the median normalized color at each radii is shown. For both star-forming and quiescent galaxies we see a large amount of scatter in the distribution of profiles. This scatter increases at small radii, $r < 2\, \rm kpc$, where some galaxies are much redder and others are significantly bluer in the center compared to the outskirts. At larger radii the scatter also increases, this is likely due to the lower S/N in the outskirts of galaxies.

We quantify the slope, or strength, of color gradients for our galaxy sample. The slope of the color profile, $d (g-r) / d \log r$ is measured simply using the change in a galaxy's median color between two radii. We choose two sets of radii: a constant physical separation of 2 kpc to 8 kpc and between $0.5\ r_{\rm light}$ and $2\ r_{\rm light}$, based on the F160W half light radius of each galaxy. Figure~\ref{fig:dc_dr} displays the median color gradient as a function of stellar mass and redshift for star-forming and quiescent galaxies in our sample.

The trends qualitatively observed above are confirmed in Figure~\ref{fig:all_col_prof}. Quiescent galaxies have negative but shallower color gradients ($d (g-r) / d \log r \sim -0.15$) compared to star-forming galaxies, that do not correlate strongly with either stellar mass or redshift. Whereas for star-forming galaxies the strength of the gradients depends strongly on redshift. The median slope changes rapidly from $-0.3$ to $-0.5$ around $z = 1.5$. The median color gradient as a function of stellar mass appears to be different for our two estimates (one based on the physical radius and the other based on the half-light radius). This is likely due to the steepening of the size mass relation at high masses \citep{Mowla2019,Mowla2019b}.

\begin{figure*}
    \centering
    \fig{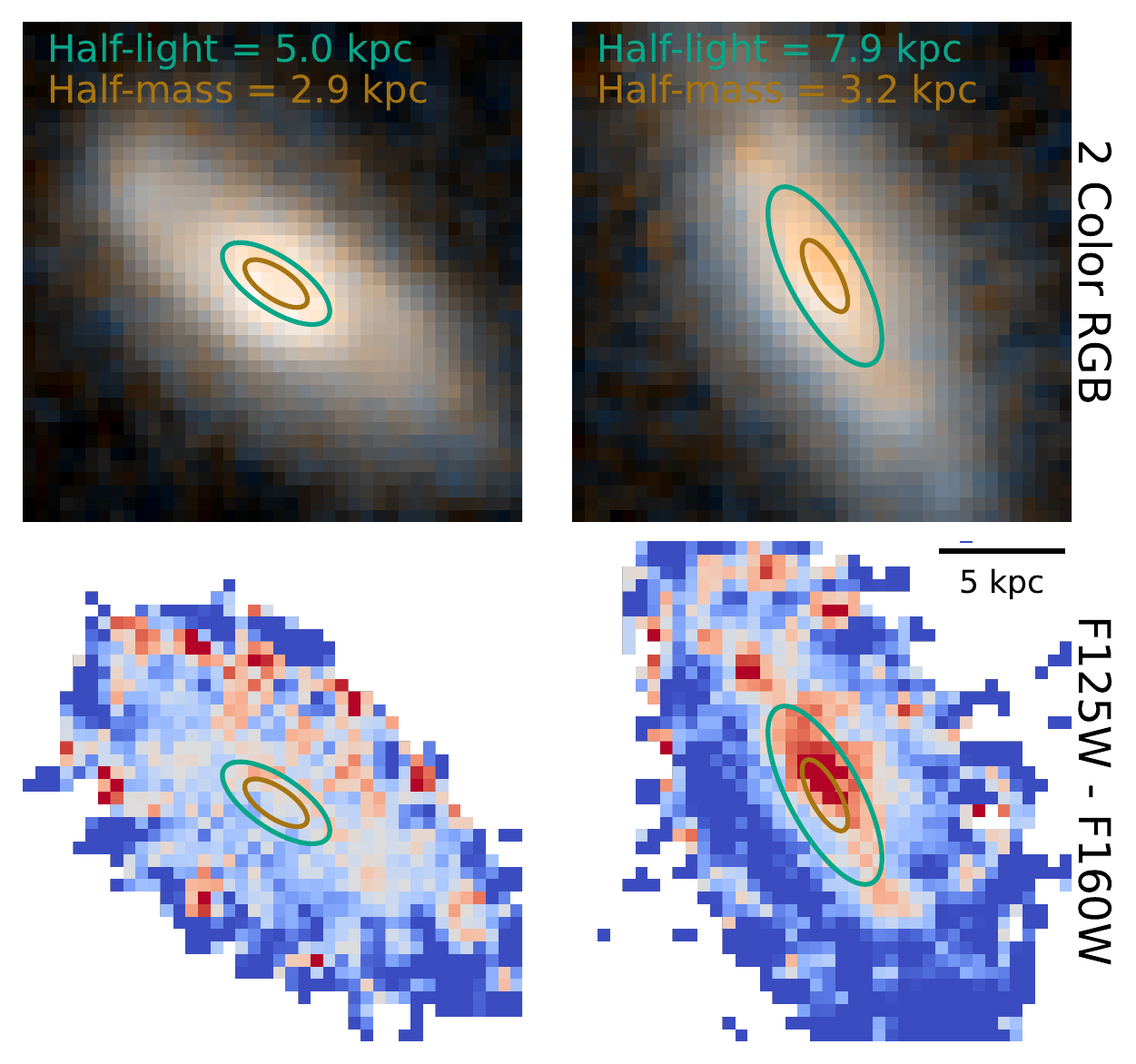}{0.45\textwidth}{\textbf{(a)} $z\approx1.2$, $M_*\approx 5 \times 10^{10}$, $q \approx 0.35$}
    \fig{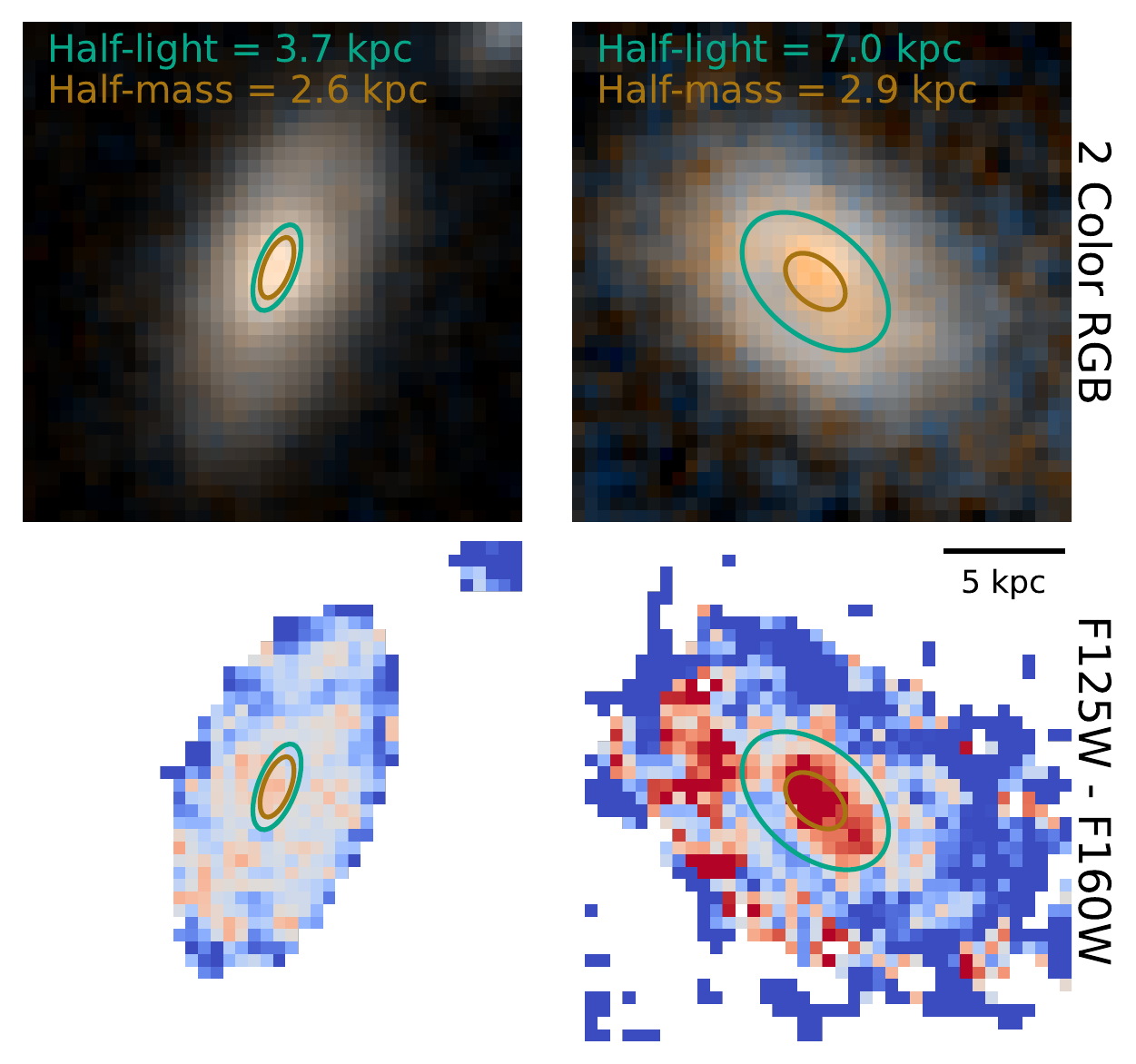}{0.45\textwidth}{\textbf{(b)} $z\approx1$, $M_*\approx 7 \times 10^{10}$, $q \approx 0.5$}

    \caption{Two sets of two galaxies which are matched based on $z,\, \log\,M_*/M_\odot, q_{\rm light}$ and $r_{\rm mass}$ but have different $r_{\rm light}$. We show color renditions of the galaxies based on the F160W and F125W images (top) and their observed color maps (bottom), obtained by dividing the images. Cyan and gold ellipses display the half-light and half-mass radius for each galaxy. The presence of color gradients affects the relationship between $r_{\rm mass}$ and $r_{\rm light}$.}
    \label{fig:r_match_examp}
\end{figure*}

Although with only two wavelength bands it is difficult to investigate the physical cause of color gradients we attempt to gain some insight in Figure~\ref{fig:dcdr_UVJ} by calculating the median color gradients across the rest-frame $U-V$ vs $V-J$ plane. Here we focus on the gradient measured between 0.5 and 2 $r_{\rm light}$, but note that the results are similar if we instead use the definition based on physical radii. There is a clear trend where the color gradients are stronger moving from bottom left to the top right in the UVJ plane. This resembles the dust track, which follows the effect of increasing dust attenuation given a standard attenuation curve ~\citep{Williams2009,Leja2019}, additionally recently interpreted as part of an evolutionary track by \citet{Suess2021}. We infer that the color gradients at $1<z<2$ likely reflect gradients in optical depth due to dust, rather than the traditional old bulges + young disks that are seen in spiral galaxies in the local Universe. We note that the regular trends in Fig.\ \ref{fig:dcdr_UVJ} provide some evidence that our measurements are robust; if they were driven by noise or systematic errors then we would not expect to see systematic trends with the (unrelated, from a measurement perspective) integrated rest-frame $U-V$ and $V-J$ colors.

\section{The relationship between mass and light weighted radii}
\label{sec:hl_hm}

In this section we investigate how color gradients, which trace (or in our case directly map to) \MLR{} gradients, interact with the light profile to produce the difference between half-light and half-mass radii. To begin we examine galaxies with similar properties, but different ratios of $r_{\rm mass}/ r_{\rm light}$. Specifically we pick star-forming galaxies with similar (within $10\%$) stellar mass, redshift, observed axis ratio and $r_{\rm mass}$, but differing $r_{\rm light}$. Figure ~\ref{fig:r_match_examp} displays the observed images and colors for two sets of galaxies. As expected from our derivation of $r_{\rm mass}$, the major difference between the galaxies is the strength of the color gradient. The centers of the galaxies with  $r_{\rm mass} \ll r_{\rm light}$ are significantly redder than their outskirts, whereas the observed color of galaxies with $r_{\rm mass} \approx r_{\rm light}$ shows little variation as a function of radius.
\begin{figure*}
    \centering
    \includegraphics[width = \textwidth]{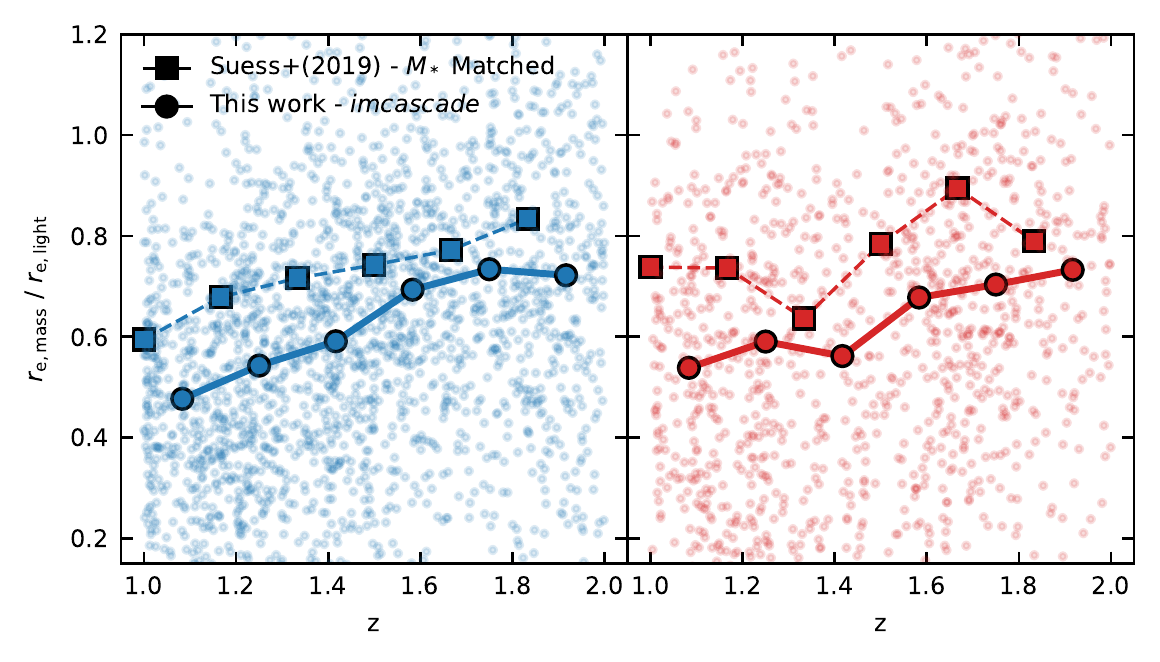}
    \caption{The evolution of the ratio between $r_{\rm mass}$ and $r_{\rm light}$. Star-forming galaxies are shown in blue in the left panel and quiescent galaxies in red in the right panel. In each panel the points show individual galaxies while the line shows the median as a function of redshift.  There is significant scatter in $r_{\rm mass}\, /\, r_{\rm light}$ but for both star-forming and quiescent galaxies we see a general trend where this ratio decreases at lower redshift. Also shown are results from \citet{Suess2019}, where we have excluded galaxies in their sample below $\log\ M_*/M_\odot = 10.25$ to match our sample. Our results qualitatively agree with those of \citet{Suess2019}. }
    \label{fig:rem_rel}
\end{figure*}

Figure~\ref{fig:rem_rel} displays the redshift evolution of $r_{\rm mass}/r_{\rm light}$ for our entire sample of galaxies. The median evolution is shown along with the distribution of individual galaxies in our sample. We also show results using measurements from \citet{Suess2019}, where we exclude galaxies below our mass limit ($\log M_* > 10.25$) to enable a direct apples-to-apples comparison.

We find similar results to \citet{Suess2019}, with a smooth decline in the average ratio of $r_{\rm mass}/r_{\rm light}$ by about a factor of $\sim 1.5$ from $z=2$ to $z=1$ for both star-forming and quiescent galaxies. The median ratio we derive in this study is roughly $10\%$ lower than that of \citet{Suess2019} at all redshifts for both star-forming and quiescent galaxies, consistent with our measurements of the half-mass radius of galaxies being slightly lower on average (see Appendix~\ref{sec:meth_comp} for a full comparison).

\begin{figure}
    \centering
    \includegraphics[width = \columnwidth]{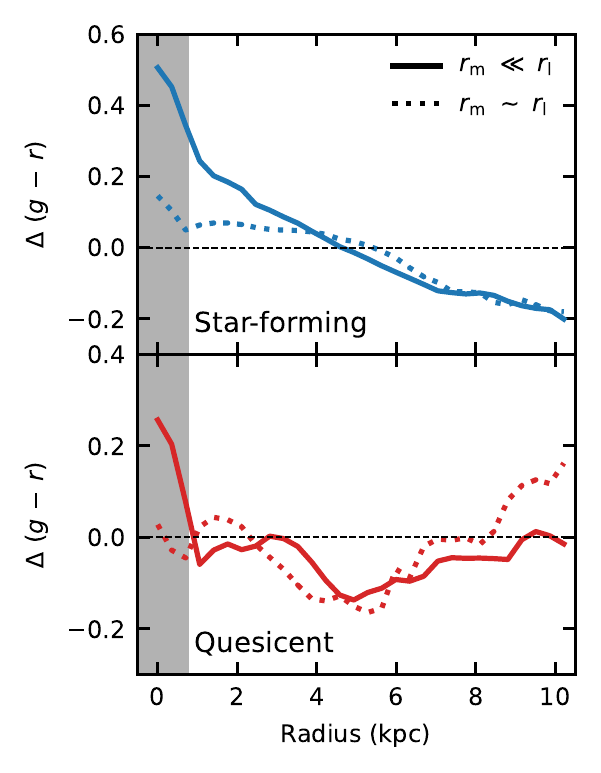}
    \caption{Comparing the properties of galaxies in the upper and lower third of $r_m\ / \ r_l$, denoted $r_{\rm m}\sim r_{\rm l}$ and $r_{\rm m} \ll r_{\rm l} $ respectively. These populations only contain galaxies with $1.4<z<1.6$ and $10.4 < \log\, M_* < 10.6$. As with previous figures the color profiles are normalized by each galaxy's total color and the light profile is normalized by its total flux. Star-forming galaxies are shown by the blue line and quiescent galaxies by the red line. There is a  clear difference between the two populations in the presence of color gradients.}
    \label{fig:rm_rl_ex}
\end{figure}

\begin{figure}
    \centering
    \includegraphics[width = \columnwidth]{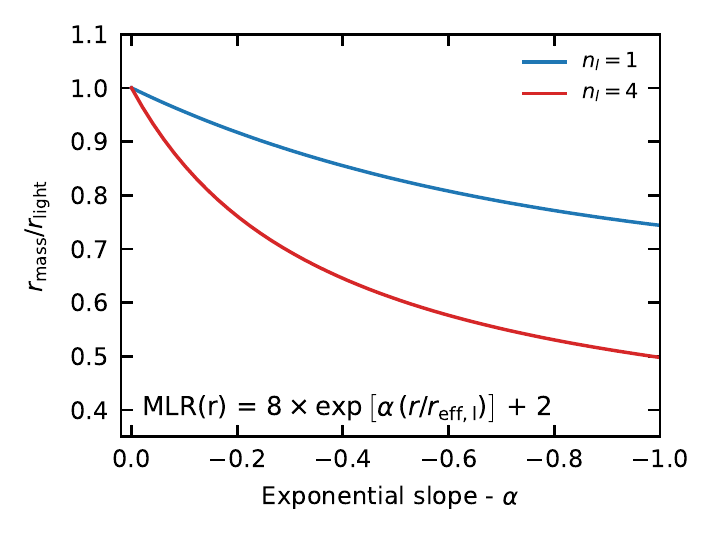}
    \caption{Results of a simple experiment of applying a varying \MLR{} gradient to two different Sersic light profiles. The resulting ratio of $r_{\rm mass}\, / r_{\rm light}$ is shown as a function of the steepness of the \MLR{} profile, parameterized by $\alpha$. It is observed that at any value of $\alpha$ the value of $r_{\rm mass}\, / r_{\rm light}$ is smaller for the more centrally concentrated $n=4$ profile compared to that with $n=1$. This explains why quiescent galaxies can have weaker color gradients than star-forming galaxies but similar values of $r_{\rm mass}\, / r_{\rm light}$.}
    \label{fig:hmr_model}
\end{figure}

Looking at the distribution of individual galaxies we see a significant amount of scatter around the median evolution. For both star-forming and quiescent galaxies, the observed scatter of $r_{\rm mass}/r_{\rm light}$ around the median, estimated using the bi-weight scale, is $0.3$. This is much larger than the median $1-\sigma$ observational uncertainty of $r_{\rm mass}/r_{\rm light}$, which is $0.09$ for quiescent galaxies and $0.05$ for star-forming galaxies, suggesting that a large amount observed scatter is driven by intrinsic variation within the galaxy population (reflecting the scatter in color profiles discussed earlier). 

\begin{figure*}
    \centering
    \includegraphics[width = \textwidth]{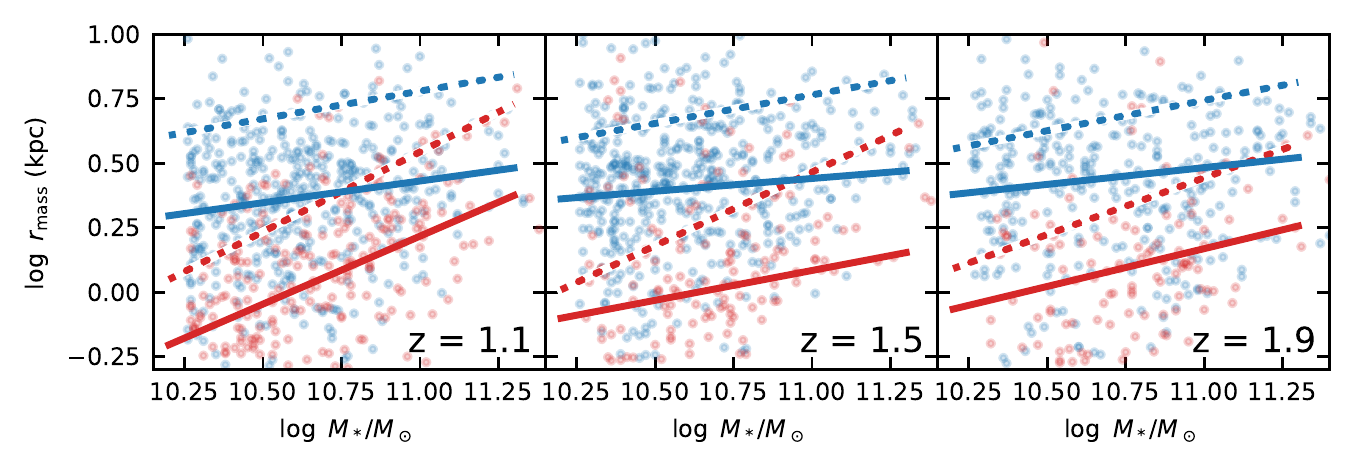}
    \caption{The $r_{\rm mass}$ - stellar mass relation for star-forming and quiescent galaxies in three redshifts bins. The distribution of star-forming galaxies is shown by blue points, with red points for quiescent galaxies. Solid lines show the best fit power-law relations in each redshift bin, described in Table~\ref{tab:size_mass_fit}. Dotted lines show the best fit relations to the  $r_{\rm light}$ - stellar mass relation for comparison. }
    \label{fig:size_mass}
\end{figure*}

To help understand the differences between galaxies with $r_{\rm mass} \approx r_{\rm light}$ and those with $r_{\rm mass} \ll r_{\rm light}$ we plot color profiles of these two populations. Galaxies are selected with $1.4<z<1.6$ and $10.4< \log\, M_*/M_\odot < 10.6$. We then select the top and bottom third of galaxies ranked by radii ratio $r_{\rm mass} / r_{\rm light}$. The top and bottom third for star-forming (quiescent) galaxies has a radii ratio of 0.98 and 0.33 (0.99 and 0.42) respectively. These two populations are denoted $r_{\rm mass}\sim r_{\rm light}$ and $r_{\rm mass} \ll r_{\rm light} $ respectively. Again there is a clear difference between the two populations in the color profile. For star-forming galaxies with $r_{\rm mass} \ll r_{\rm light}$, the color of the center is $\sim 0.75$ mag redder than the outskirts, where for the population with $r_{\rm mass} \approx r_{\rm light}$, the color difference is only $\sim 0.2$ mag. For quiescent galaxies the difference is only apparent in the inner 1 kpc. Unfortunately we cannot resolve much of this behavior due to the PSF.

Another interesting observation is that the average ratio of $r_{\rm mass} / r_{\rm light}$ for star-forming and quiescent galaxies looks very similar, yet their color gradients (shown in Figure ~\ref{fig:all_col_prof}) appear very different. These two populations have very different light profiles on average. Quiescent galaxies tend to have more centrally concentrated (i.e., high Sersic index) profiles compared the star-forming galaxies which often show exponential-like profiles \citep{Mowla2019b}.

To investigate the effect of the light profile on the conversion between $r_{\rm mass} $ and $ r_{\rm light}$ we set up an experiment based on one dimensional profiles. We start with two Sersic profiles to represent the light profiles of galaxies: $n=1$ to mimic star-forming galaxies and $n=4$ to mimic quiescent galaxies. We then apply a mass-to-light ratio gradient based on an exponential decline,
\begin{equation}
    M/L\, (r) = 8 \times \exp\left[ \alpha\  (r/r_{\rm eff,l}) \right] + 2.
\end{equation}
We choose this form to roughly match realistic \MLR{} profiles, however the following conclusions are not sensitive to the details of the chosen profile.

The effect of changing the exponential slope $\alpha$ on the ratio $r_{\rm mass} / r_{\rm light}$ is shown in Figure~\ref{fig:hmr_model}. For both profiles, the half-mass radius decreases as the slope of the \MLR{} gradient gets steeper, as expected. At a fixed value of $\alpha$ the $n=4$ profile has a smaller half-mass radius compared to the $n=1$ profile. The $n=4$ profile is more centrally concentrated and since applying the \MLR{} is a multiplicative effect, the mass profile is even more centrally concentrated compared to an exponential profile. This offers an explanation as to why star-forming and quiescent galaxies can have differing color gradients yet similar evolution of  $r_{\rm mass} / r_{\rm light}$.

\begin{figure}
    \centering
    \includegraphics[width = \columnwidth]{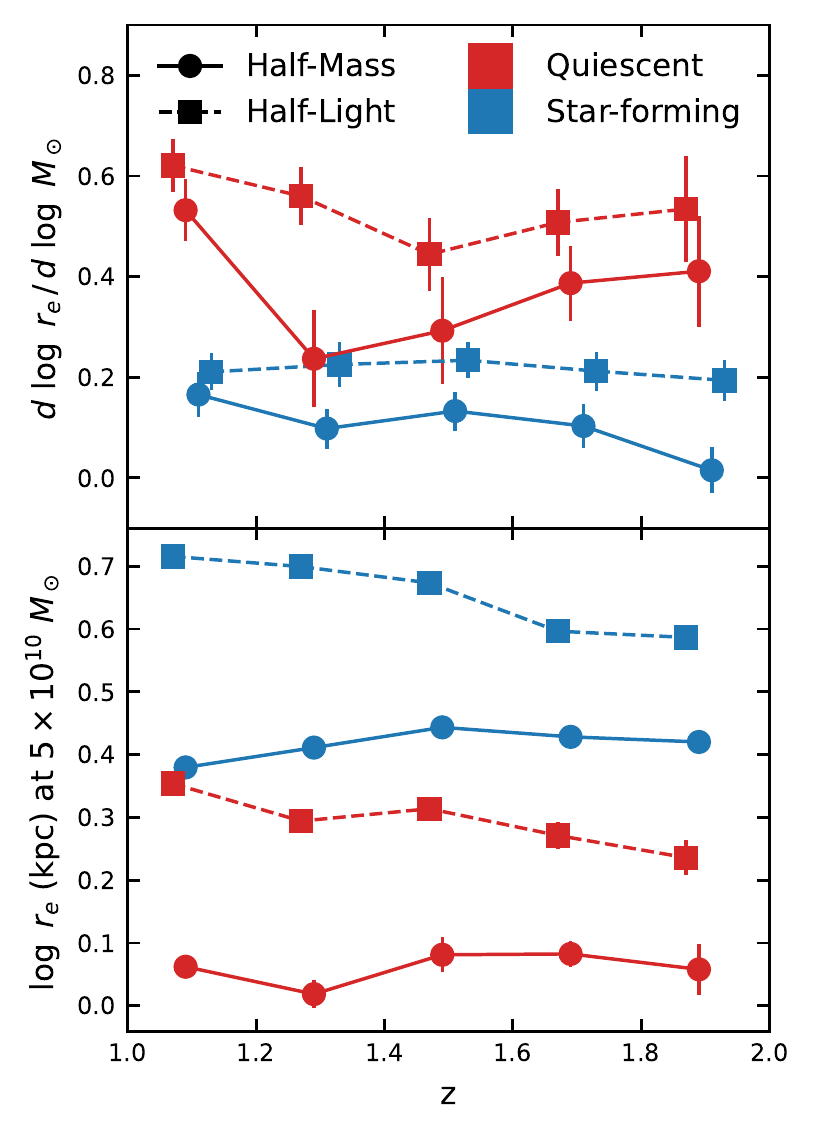}
    \caption{The evolution of best-fit parameters to the $r_{\rm mass}$ - stellar mass relation. The evolution of the slope (top panel) and normalization (bottom panel) is shown for both star-forming and quiescent galaxies. Also shown for comparison are the best fit parameters of the  $r_{\rm light}$ - stellar mass relation. The slopes of the $r_{\rm mass}$ relations are shallower than their $r_{\rm light}$ counterparts and we see much less evolution in the normalization.} 
    \label{fig:size_mass_params}
\end{figure}

\section{The Galaxy Half-Mass size - Stellar Mass Relation}
\label{sec:size-mass}

In this section we study the $r_{\rm mass}$ - $M_*$ relation, comparing it to the well-studied $r_{\rm light}$ - $M_*$ relation. The $r_{\rm mass}$ - stellar mass relation for galaxies in our sample at three different redshift bins is shown in Figure~\ref{fig:size_mass}. To investigate the evolution of the relation and compare to previous results we fit a power-law to the $r_{\rm mass}$ - stellar mass relation. This is a common parameterization of the size mass-relation ~\citep{vanderWel2014,Mowla2019b}. Other studies have used a broken power law or other functions with a changing slope at low mass to fit the size-mass relation \citep{Shen2003,Lange2015,Mowla2019,Mosleh2020}, however since our sample only contains galaxies with $\log\, M_*/M_\odot > 10.25$ we cannot constrain the low-mass slope.

The relations for star-forming and quiescent galaxies are fit separately in each redshift bin. Galaxies with median $r_{\rm mass}$ less than 1 pixel (equal to 0.06 arcsec or $\sim 0.5$ kpc at $z=1.5$) are excluded from the fit. We fit the following functional form,
\begin{equation}
    \log_{10} (r_{\rm mass}/{\rm kpc}) = s\, \log_{10} \left( M_*/ (5\times 10^{10} M_\odot) \right) + b,
\end{equation}
with $s$ (the slope) and $b$ (the intercept or average size at $ M_*=  (5\times 10^{10} M_\odot$) are free parameters. The fit is performed by minimizing the least squares residuals utilizing five rounds of $\sigma$-clipping with a $4\sigma$ threshold. With 500 bootstrap samples of each population we take the median values of $s$ and $b$ as the best-fit value and the uncertainty as half of the distance between the 16th and 84th percentile. The best fit parameters for each population are shown in Table~\ref{tab:size_mass_fit}. We compare our best fit parameters for the $r_{\rm light}-M_*$ relation to previous works in Appendix~\ref{sec:size_mass_comp} and show that they agree well.

\begin{table}[]
    \centering
    \caption{The best fit parameters to the half-mass size-stellar mass relation for star-forming and quiescent galaxies of the form: $\log_{10} (r_{\rm mass}/{\rm kpc}) = s\, \log_{10} \left( M_*/ (5\times 10^{10} M_\odot) \right) + b$}

    \begin{tabular}{c|c c | c c}
        $z$ & \multicolumn{2}{|c|}{Star-forming} & \multicolumn{2}{|c}{Quiescent} \\ \hline
         & $s$ & $b$ & $s$ & $b$ \\ \hline \hline
        $1.1$ & $0.17 \pm 0.04 $ &$ 0.38 \pm 0.01 $& $0.53 \pm 0.06$ & $0.06 \pm 0.02$ \\
        $1.3$ & $0.09 \pm 0.04 $ &$ 0.41 \pm 0.01 $& $0.22 \pm 0.09$ & $0.02 \pm 0.02$ \\
        $1.5$ & $0.14 \pm 0.04 $ &$ 0.44 \pm 0.01 $& $0.30 \pm 0.09$ & $0.08 \pm 0.03$ \\
        $1.7$ & $0.10 \pm 0.05 $ &$ 0.43 \pm 0.01 $& $0.40 \pm 0.06$ & $0.08 \pm 0.02$ \\
        $1.9$ & $0.01 \pm 0.04 $ &$ 0.42 \pm 0.01 $& $0.42 \pm 0.11$ & $0.06 \pm 0.04$ \\ \hline
    \end{tabular}
    \label{tab:size_mass_fit}
\end{table}

The best fit lines for the $r_{\rm light}$ and $r_{\rm mass}$ relations for quiescent and star-forming galaxies are shown in each panel of Figure~\ref{fig:size_mass}. Consistent with the results in Section~\ref{sec:hl_hm} we find that the half-mass radii of galaxies are smaller at all redshifts and masses compared to the half-light radii. We observe that the slope of the $r_{\rm mass}-M_*$ relation appears to be shallower for both star-forming and quiescent galaxies and the relations appear to evolve less with redshift.

\begin{figure}
    \centering
    \includegraphics[width = \columnwidth]{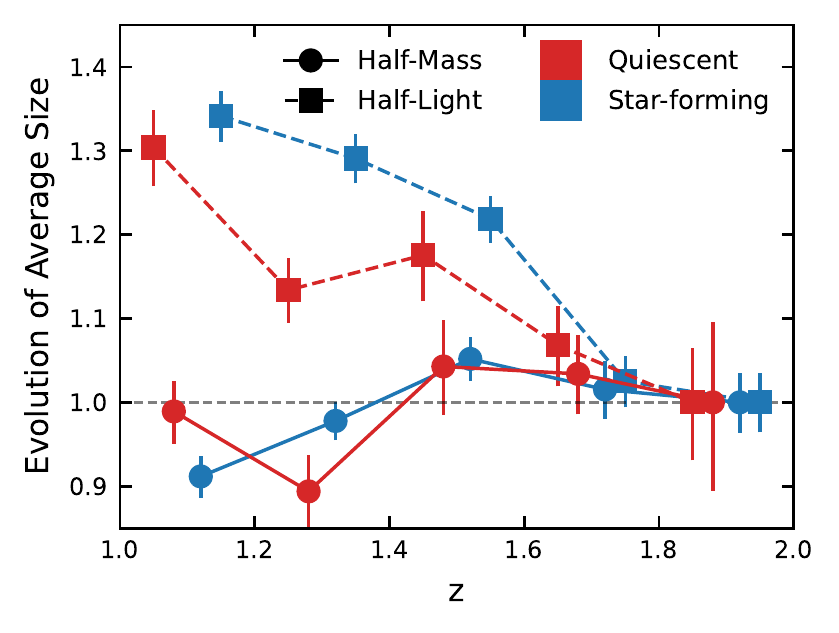}
    \caption{The evolution of the normalization of the size-mass relation, $b$, is highlighted. For each curve we normalize the growth to the highest redshift bin at $z=1.9\, \pm\, 0.1$. Each curve is offset slightly on the x axis for clarity. The average $r_{\rm mass}$ does not evolve (or even gets smaller) for both star-forming and quiescent galaxies, whereas $r_{\rm light}$ grows by roughly 30\,\%.}
    \label{fig:med_size_evo}
\end{figure}

The best fit parameters for the $r_{\rm mass}\ -\ M_*$ relation are compared those of the $r_{\rm light}\ -\ M_*$ relation as a function of redshift in Figure~\ref{fig:size_mass_params}. The slope of the half-mass relation is lower on average compared to the half-light relation. For quiescent galaxies, the half-mass slope is $\approx 0.4$, compared to  $\approx 0.6$ for the half-light relation. The slope for star-forming galaxies of the half-mass slope roughly matches that for half-light sizes at $z=1$ but decreases to higher redshift. At $z=2$, the star-forming $r_{\rm mass}\ -\ M_*$ is consistent with being flat. The normalization evolves similarly for star-forming and quiescent galaxies. The best fit value for $b$ remains constant over the redshift range of our sample while the average $r_{\rm light}$ value increases steadily over the redshift range for both populations of galaxies.

The difference in the evolution of the average size between half-mass and half-light radii is highlighted in Figure~\ref{fig:med_size_evo}. In this figure we normalize the evolution of $b$ to that of the highest redshift bin ($z = 1.9 \pm 0.1$). This highlights the striking difference between $r_{\rm mass}$ and $r_{\rm light}$. The ratio between the average r$_{\rm light}$  at z=1 to z = 2 is by $35\pm 3\%$ ($31\pm 4\%$) for star-forming (quiescent). However, for $r_{\rm mass}$ this ratio is much more consistent: $-2 \pm 4 \%$ and $9\pm 2\%$ for quiescent and star-forming galaxies respectively. This result is consistent with \citet{Suess2019} and \citet{Mosleh2020} who find little evolution in the average half-mass size of galaxies between $1<z<2.5$.

\section{Discussion} \label{sec:disc}
\subsection{Comparison to previous work}

In this study we use \imc~to measure color gradients for galaxies at $1<z<2$. \imc~utilizes the MGE formalism to flexibly measure surface-brightness and color profiles. Half-mass radii are measured by converting the color profile into an \MLR{} using an empirical relation. This is not the first such analysis, and here we discuss differences and similarities with previous work. Previous studies have relied on parametric representation for either the surface-brightness profile of the \MLR{} profile. \citet{Suess2019} test three methods, but the authors favor a forward modeling approach which uses a Sersic function for the light profile and a power-law to represent the \MLR{} profile, which is compared to the observed \MLR{} profile, measured by SED fitting in consecutive annuli. \citet{Mosleh2020} perform SED fitting directly to observed pixels and then use a Sersic profile to model the resulting stellar mass image.

The flexibility of the MGE method is the strength of our procedure. The light and color profiles, which are likely complex and multi-component, are measured directly without prescribing a parameterization. In Figure~\ref{fig:all_col_prof} we find non-monotonic and multi-component color profiles. The other side of this coin is that we are using a simple, empirically derived relation, to convert from observed color to \MLR{}.  This is subject to systematic uncertainties and large scatter, as shown in Figure~\ref{fig:col_fit}. The SED fitting techniques employed by \citet{Suess2019} and \citet{Mosleh2020} likely result in a lower scatter estimate of the observed \MLR{}. These are based on flux measurements from several bands which is not possible in this work. Given these major differences in the procedures used it is heartening to see that measurements all three methods generally agree. 

An additional concern beyond measuring the \MLR{} profile is its interaction with the light profile. As we show in Figure~\ref{fig:hmr_model} and is also discussed in \citet{Bernardi2022}, the structure of the galaxy also impacts the measurement of $r_{\rm mass}$. Due to their central light concentration, for galaxies with a larger observed Sersic index, the same \MLR{} gradient produces a much lower ratio of $r_{\rm mass} / r_{\rm light}$. Therefore to accurate measure $r_{\rm mass}$ one needs to accurately measure both the \MLR{} and light profiles. 

Given the subtleties involved in measuring $r_{\rm mass}$ it is encouraging that our measurements generally agree with the previous results of ~\citet{Suess2019,Mosleh2020}. A More detailed comparisons of these measurements is shown in Appendix~\ref{sec:meth_comp}. The half-mass radii agree to within 10\% between the three studies and the ratio of $r_{\rm mass}/r_{\rm light}$ 20\% on average. However the observed scatter, of roughly $0.2$ dex, is larger than the combination of quoted uncertainties indicating systematic differences on the level of individual galaxies.

\subsection{Physical drivers of the color gradients}

The color gradients measured for quiescent galaxies in our sample match well with observations of early-type galaxies (ETG) in the local universe. \citet{Wu2005}, \citet{LaBarbera2009} and \citet{Tortora2011} all find negative color gradients on average for ETGs in the local universe with mass greater than $\log M_\odot/M_* \gtrsim 10.5$. Each study finds mildly negative color gradients, $d(g-r)/d\log R \sim 0.1$, similar to our study, for their galaxy samples. Utilizing SED fitting of optical and NIR bands, all three studies attribute metallicity gradients as the dominant cause, as opposed to stellar age gradients, the other possibility considered. If the quiescent galaxy populations at $z=1-2$ are related to those in the local universe, the fact that we measure similar color gradients supports the conclusion that they are driven by metallicity rather than age. \citet{LaBarbera2009} finds a slight positive age gradient ($\langle\Delta \log ({\rm age})/\Delta \log R \rangle = 0.13$), which can explain the observed evolution of the half-light sizes of ETG's. In their model, ETGs form at $z\approx 2$, with younger and therefore brighter centers. The central stellar population then ages and dims, causing the half-light radius to increase. Using a simple burst SFH, this can explain the observed evolution of their half light sizes. This is consistent with our findings that the mass radii of quiescent galaxies do not evolve significantly with redshift.

An additional driver of \MLR{} gradients in quiescent galaxies is variation in the stellar initial mass function (IMF). Recent studies have suggested that the IMF is more bottom heavy in the center of some massive elliptical galaxies in the local universe \citep{Conroy2017,DominguezSanchez2019,VanDokkum2021}. This effect is essentially impossible to measure from  photometry alone, yet can still drastically alter \MLR{}. \citet{DominguezSanchez2019} use the \textit{MANGA} IFU survey to measure the systematic variations in the IMF of elliptical galaxies as function of radius. \citet{Bernardi2022} investigate how these variations manifest as \MLR{} gradients. The authors find that the inclusion of an IMF gradient, in coordination with the centrally concentrated light profile of elliptical galaxies, can decrease half-mass size by up to $30\%$. Such changes would be additional to the results presented here.

Previous studies of color gradients in star forming galaxies at cosmic noon suggest they are mostly due to variation in dust content. \citet{Liu2016} and \citet{Liu2017} use measured NUV and optical colors, combined with SED fitting, to infer that color gradients in galaxies along the star-forming sequence from $0.5<z<2$ are predominantly due to variation in dust extinction. This generically produces negative color gradients as the centers of galaxies have a dust higher optical depth than the outskirts. \citet{Wang2017} measure radial gradients in the UVI color plane, similar to the more common UVJ variant, and find gradients within galaxies are almost parallel to the expected dust extinction vector. These findings are supported by studies of the Balmer decrement, that have found that dust extinction is higher in the centers of galaxies \citep{Nelson2016b} and recent studies showing the FIR sizes, which traces dust mass, is typically smaller than optical sizes\citep{Tadaki2020,Gomez-Guijarro2022}.  

Our study provides additional evidence that color gradients in star-forming galaxies are caused by dust extinction. In Figure~\ref{fig:dcdr_UVJ} we find the strongest color gradients in the region of the UVJ plane where dust extinction is expected to be the largest. We note that this trend has also been interpreted as an evolutionary track, in \citet{Suess2021}. In the local universe \citet{Kennedy2015} find that for star-forming galaxies almost all the variation in Sersic index and half-light radius with wavelength can be attributed to changes in dust opacity and inclination angle. Similarly \citet{Patel2012} show that at $0.6 < z < 0.9$ a galaxy's location along the dust vector in the UVJ plane is strongly correlated with inclination angle. This is again suggestive that dust opacity and geometry plays a large role in shaping the morphology and color gradients of star-forming galaxies.

\begin{figure}
    \centering
    \includegraphics[width = \columnwidth]{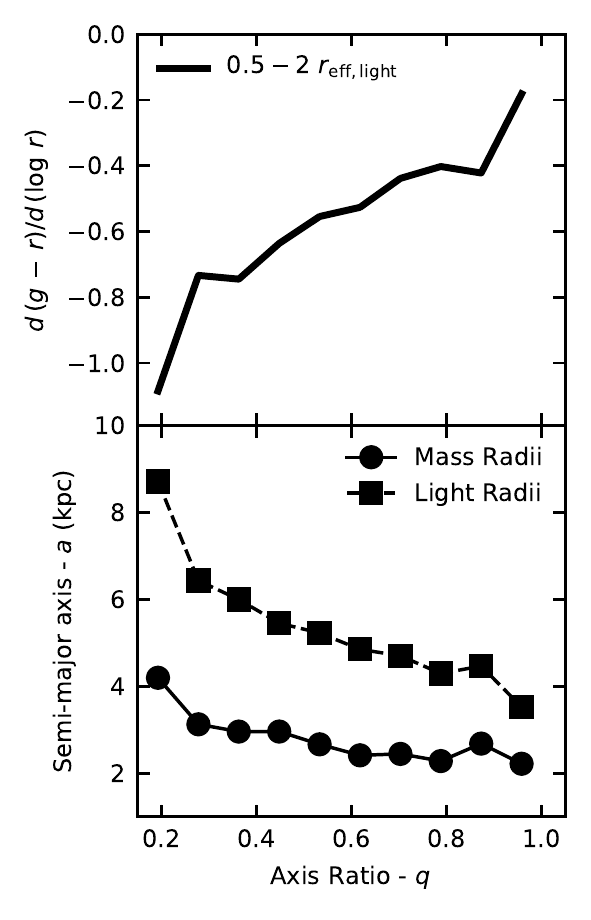}
    \caption{We confirm the results of \citet{Mowla2019c}, finding an inclination bias in the half-light sizes of galaxies. Similar to their study we focus on star-forming galaxies at $1<z<1.25$ and $10.5 < \log\, M_*/M_\odot < 11$. We see a clear correlation between axis ratio and $r_{\rm light}$. We also plot the median color gradient and half-mass size as a function of $q$. There is a clear correlation between color gradient and $q$ whereas $r_{\rm mass}$ does not depend on the axis ratio. These results are consistent with a scenario where the stars and dust are present in a (relatively) thin disk, leading to a bias in $r_{\rm light}$ for certain viewing angles.}
    \label{fig:a_q}
\end{figure}

Related to this is the issue of an inclination bias in the size-mass plane, as suggested by \citet{Mowla2019c}. The authors investigate a discrepancy between the observed line width and that expected by simple viral estimator for three compact star-forming galaxies. They find the most plausible explanation is that these galaxies are mostly face on disks. Additionally they show a clear trend between $r_{\rm light}$ and axis ratio for these galaxies, suggesting that compact star-forming galaxies may simply be a face-on projection rather than a separate class. In Figure~\ref{fig:a_q} we recreate this relationship using both the light and mass semi-major axis. We find a similar correlation to \citet{Mowla2019c} between the semi-major axis and axis ratio. For mass radii there is much less correlation. Additionally we show that there is a clear correlation, where galaxies with lower axis ratios have steeper color gradients. A simple explanation is that the stars and dust in these galaxies live in (relatively) thin disks. Therefore the dust attenuation is higher when looking edge-on. Similar to \citet{Mowla2019c},we find an inclination bias when using $r_{\rm light}$. Accounting for color gradients and measuring half-mass sizes, appears to mostly alleviate this bias.

\begin{figure*}
    \centering
    \includegraphics[width = 0.95\textwidth]{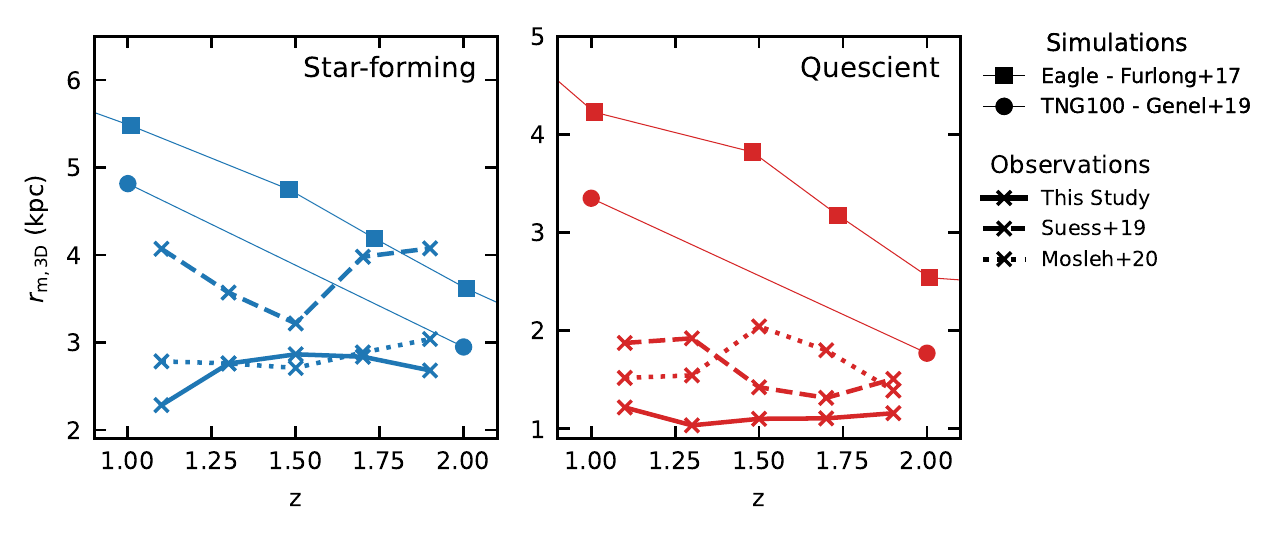}
    \caption{The sizes of galaxies at fixed mass from $z=\,1\,-\,2$, comparing observations with predictions from hydrodynamical simulations. We show the median 3D half-mass sizes within a fixed stellar mass range of $\log M_*/M_\odot = 10.5-11$. For the TNG and Eagle simulations the measurements are taken from \citet{Genel2018} and \citet{Furlong2017}, see text for details. The observed 3D radii are calculated using the circularized half-mass radii as prescribed in \citet{vandeVen2021}. The simulations and observations agree fairly well at $z\sim 2$. However, the simulations predict significant change in the sizes of galaxies as a function of redshift  whereas the observations suggest there is very little variation with redshift in either star-forming or quiescent galaxies.}
    \label{fig:sim_comp}
\end{figure*}

\subsection{Comparison to simulations}

A benefit of using half-mass radii is that they offer a relatively direct comparison to size predictions in numerical galaxy formation simulations. In order to properly compare simulations to observations, it is common to forward model the simulation output and apply observational techniques. This is a non-trivial process that involves expensive radiative transfer calculations to properly account for dust extinction. Additionally simulations do not fully resolve the interstellar medium of galaxies and many do not self-consistently track the production and destruction of dust leading to additional uncertainties. This has been implemented successfully by several studies \citep{Price2017,RodriguezGomez2019, Bignone2020,Parsotan2021,deGraaff2022} and is an important consistency check. However the process of creating mock observations can make it difficult to investigate the underlying causes of discrepancies. Using half-mass radii provides a more direct way to compare observations to simulations.

The observed evolution of $r_{\rm mass}$ is compared to results from hydrodynamical simulations of galaxies in Figure~\ref{fig:sim_comp}. For observations we use the results from this study along with catalogs published in~\citet{Suess2019} and~\citet{Mosleh2020}. To directly compare to simulations we convert the observed (projected) half-mass radius to a 3D radius based on the findings of~\citet{vandeVen2021}. The authors use analytic calculations to investigate the mapping between projected and 3D radii. They find that $1.3\times r_{\rm eff,circ}$, with $r_{\rm eff,circ}$ the circularized half-mass (or light) radius, is an unbiased estimator of $r_{\rm med}$, the 3D effective radius (with a scatter of $\approx 30\%$ from galaxy to galaxy). As we are analyzing the evolution of the median size in this figure we use this to convert from observed radii to 3D radii, which we directly compare to simulations.

We use previously published results from the EAGLE \citep{Crain2015,Schaye2015,Furlong2017} and TNG-100 \citep{Nelson2018,Pillepich2018,Marinacci2018,Naiman2018,Springel2018, Genel2018} simulations. For EAGLE, we plot the evolution of the 3D half-mass sizes of galaxies taken directly from Figure 3 in~\citet{Furlong2017} for active and passive galaxies (which we plot as star-forming and quiescent respectively). For TNG-100  we take the measurements of the median 3D half-mass sizes of galaxies at $z=1$ and $z=2$ from Figure 1 in \citet{Genel2018}. We interpolate these median relations to $\log M_*/M_\odot = 10.75$ for main-sequence and quiescent galaxies, which we plot on Figure~\ref{fig:sim_comp} as star-forming and quiescent respectively. We note that both \citet{Furlong2017} and \citet{Genel2018} directly compare 3D mass radii, or mock observations without accounting for dust, to observed half-light radii which we argue is not an apples-to-apples comparison and should be interpreted with caution.

The measurements from simulations and observations appear consistent at $z=2$, however their evolution is qualitatively different. As seen in~\citet{Suess2019},~\citet{Mosleh2020} and we show in Section~\ref{sec:size-mass}, the observed half-mass radii for both star-forming and quiescent galaxies are not significantly different over the redshift range $z=1-2$. By contrast, for both the EAGLE and TNG-100 simulations the mean size of galaxies changes $50\%-75\%$ over the same epoch. This growth in radii of simulated galaxies roughly matches that observed in half-light radii ~\citep{vanderWel2014,Mowla2019}, yet is starkly different when an apples-to-apples comparison to $r_{\rm mass}$ is made.

In Table~\ref{tab:size_mass_fit}, we provide the results of power-law fits to the $r_{\rm mass}-M_*$ relation. These measurements agree well with previous studies~\citep{Suess2019,Mosleh2020}\footnote{We note that in \citet{Mosleh2020}, the authors opt for a broken power law fit making comparison more difficult. However focusing on the high mass slope and normalization, our measurements agree within quoted uncertainties.}. For quiescent galaxies the relation is still steeper than the star-forming relation however the slope is when using $r_{\rm mass}$ ($d\,\log\, r_{\rm mass} /\ d\,\log\, M_* \sim 0.4$ ) is shallower than when using $r_{\rm light}$ ($d\,\log\, r_{\rm light} /\ d\,\log\, M_* \sim 0.6-0.8$ ) \citep{vanderWel2014,Mowla2019}. Combined with the lack of evolution in median size of quiescent galaxies, this seems inconsistent with the theory that quiescent galaxies grow by dry-minor mergers, which predicts a large increase in radius roughly following: $d\, \log\, r /d\,\log\, M_* \propto 2$ \citep{vanDokkum2010}. However it is possible that the time between $z=1-2$, is not long enough to see significant evolution in the population due to mergers. \citet{Suess2019c} extend the previous study down the $z\sim0$ and find the median $r_{\rm mass}$ of quiescent galaxies does evolve significantly below $z=1$, consistent with a model of growth based on minor mergers~\citep{Newman2012}. The flat slope and little evolution in the $r_{\rm mass}-M_*$ relation for star-forming galaxies, favors self-similar or spatially coherent growth. This scenario, characterized by flat sSFR profiles, has been suggested by observational studies at high redshift \citep{Liu2016,Nelson2016} and by hydrodynamical simulations of galaxies \citep{Nelson2021}.

\section{Summary and Conclusions} \label{sec:conc}

In this work we measure color gradients and half-mass radii of galaxies at $1<z<2$. We utilize \imc, a Bayesian implementation of MGE representation, to flexibly measure the surface brightness and color profiles of galaxies in our sample. By using an empirical relationship between color and \MLR{}, we measure half-mass radii for galaxies as well. We characterize the $r_{\rm mass}-M_*$ relationship comparing it to the well studied $r_{\rm light}-M_*$ relationship.

We typically find negative color gradients, with the slope for star-forming galaxies steeper than for quiescent galaxies. The gradients roughly match those measured in the local universe. We find significant variation in the color gradients of galaxies at a fixed $M_*$ and $z$, leading to intrinsic scatter in $r_{\rm mass}/r_{\rm light}$.

We confirm previous results by \citet{Suess2019} and show that $r_{\rm mass}/r_{\rm light}$ decreases by $\sim 50\%$ between $z=2$ and $z=1$. Even though the color gradients are shallower, quiescent galaxies have a similar ratio of $r_{\rm mass}/r_{\rm light}$, due to the steeper light profile.

By characterizing the $r_{\rm mass}-M_*$ relation, we find it is shallower compared to the $r_{\rm light}-M_*$ counterpart, and that the normalization changes less with redshift. The ratio of the average half-mass size between $z=1$ and $z=2$ is  $9\pm 2\%$ and $-2 \pm 4\%$ for quiescent galaxies and star-forming galaxies respectively. This is in stark contrast with the average half-light size which grows by $\geq 35\%$ for both populations. Our study adds to the growing consensus which indicates that half-mass radii evolve much slower than half-light radii. \citep{Suess2019,Mosleh2020} This challenges the traditional view of galaxy growth along with predictions from hydrodynamical simulations.

We emphasize here that evolution of {\em relations} is not the same as evolution of {\em individual galaxies}. Galaxies increase their mass over time, which means that a comparison at fixed mass (such as done in Fig~\ref{fig:med_size_evo}) is not equivalent to the size evolution of any individual galaxy. These relations are a consequence of the evolution of a population of galaxies. Galaxies may evolve along the size mass relation yet decoding the growth of individual galaxies is much more difficult as one must account for their growing stellar mass and the transition from star-forming to quiescent (or vice-versa). \citep[see][for a further discussion of these issues]{VanDokkum2015} Alternatively hydrodynamical simulations provide the ability to study the evolution of individual galaxies \citep[e.g.][]{Genel2018}.

We also stress that measuring the half-mass radii of faint and barely resolved galaxies is pushing current data to its limit. One needs to measure both color and light profiles accurately, while also converting to \MLR{}, in order to correctly measure the half-mass radii of galaxies. While it is encouraging that on average our results agree with both \citet{Suess2019} and \citet{Mosleh2020} within $\sim 10\%$ on average, there are still systematic uncertainties when comparing individual galaxies. With the coverage of longer wavelength and higher spatial resolution of the recently launched \textit{James Webb Space Telescope} we will be able to study the rest-frame NIR morphology of galaxies at $z>1$ for the first time. The NIR is a much more direct tracer of stellar mass, with less variation in \MLR{} compared to optical wavelengths. This will provide a definitive answer to the evolution of half-mass radii.

With JWST at high redshift, the next generation multi-band ground based surveys, like HSC-SSP \citep{Aihara2019} and the upcoming Rubin observatory\citep{Ivezic2019}, will be able to accurately measure half-mass radii at $z<1$ from multi-band photometry. This combination will map the evolution of half-mass radii from galaxies at high-redshift to the local universe. Using $r_{\rm mass}$ as a more fundamental tracer of the structure of galaxies, this will shed new light on the morphological evolution of galaxies.

\vspace{0.5mm}

\begin{acknowledgments}
\section*{Acknowledgments}
The authors would like the thank Arjen van der Wel for useful discussions and comments on the manuscript. TBM would like to thank Patricia Gruber and the Gruber foundation for their support of the work presented here. This work is based on observations taken by the 3D-HST Treasury Program (GO 12177 and 12328) with the NASA/ESA HST, which is operated by the Association of Universities for Research in Astronomy, Inc., under NASA contract NAS5-26555.
 
\vspace{0.5mm}
\software{\texttt{numpy} \citep{numpy}, \texttt{scipy} \citep{scipy}, \texttt{matplotlib} \citep{matplotlib}, \texttt{astropy} \citep{Astropy2018}, \texttt{dynesty} \citep{dynesty},  \texttt{pandas} \citep{pandas}, asdf \citep{asdf}, \texttt{SEP} \citep{SEP,SExtractor},  \texttt{imcascade} \citep{imc_zenodo} }
\end{acknowledgments}

\vspace{0.5mm}

\bibliography{col_grad,combined_bibs}
\bibliographystyle{aasjournal}

\appendix 
\section{Best fit relationships between observed color and physical quantities}
\label{sec:obs_col_rel}

\begin{table}[]
    \centering
    \caption{The best-fit values for the empirical relations used to convert from the observed color to other quantities of the form: $X = a\, (m_{\rm F125W} - m_{\rm F160W}) + b$}

    \begin{tabular}{c|c c c | c c c}
        $z$ & \multicolumn{3}{|c|}{$X = (M/L)_{\rm F160W}$} & \multicolumn{3}{|c}{$X = (g-r)_{\rm rest}$} \\ \hline
         & $a$ & $b$ & r.m.s. & $a$ & $b$ & r.m.s.  \\ \hline \hline
        $1.1$ & $2.52 $ &$ -1.19 $ & $0.33$& $1.74 $ &$ 0.07 $ & $0.23$ \\ 
        $1.3$ & $2.88 $ &$ -1.63 $ & $0.48$ &$1.49 $ &$ 0.01 $ & $0.24$ \\ 
        $1.5$ & $3.15 $ &$ -1.81 $ & $0.71$ &$1.39 $ &$ -0.02 $ & $0.21$ \\ 
        $1.7$ & $1.31 $ &$ -1.07 $ & $0.23$ &$0.96 $ &$ 0.18 $ & $0.12$ \\ 
        $1.9$ & $1.28 $ &$ -1.13 $ & $0.22$ &$0.99 $ &$ 0.14 $ & $0.21$ \\ 
        $2.1$ & $1.16 $ &$ -1.20 $ & $0.30$& $0.86 $ &$ 0.18 $ & $0.26$ \\ \hline
    \end{tabular}
    \label{tab:obs_col_param}
\end{table}

\begin{figure}
    \centering
    \includegraphics[width = 0.95\textwidth]{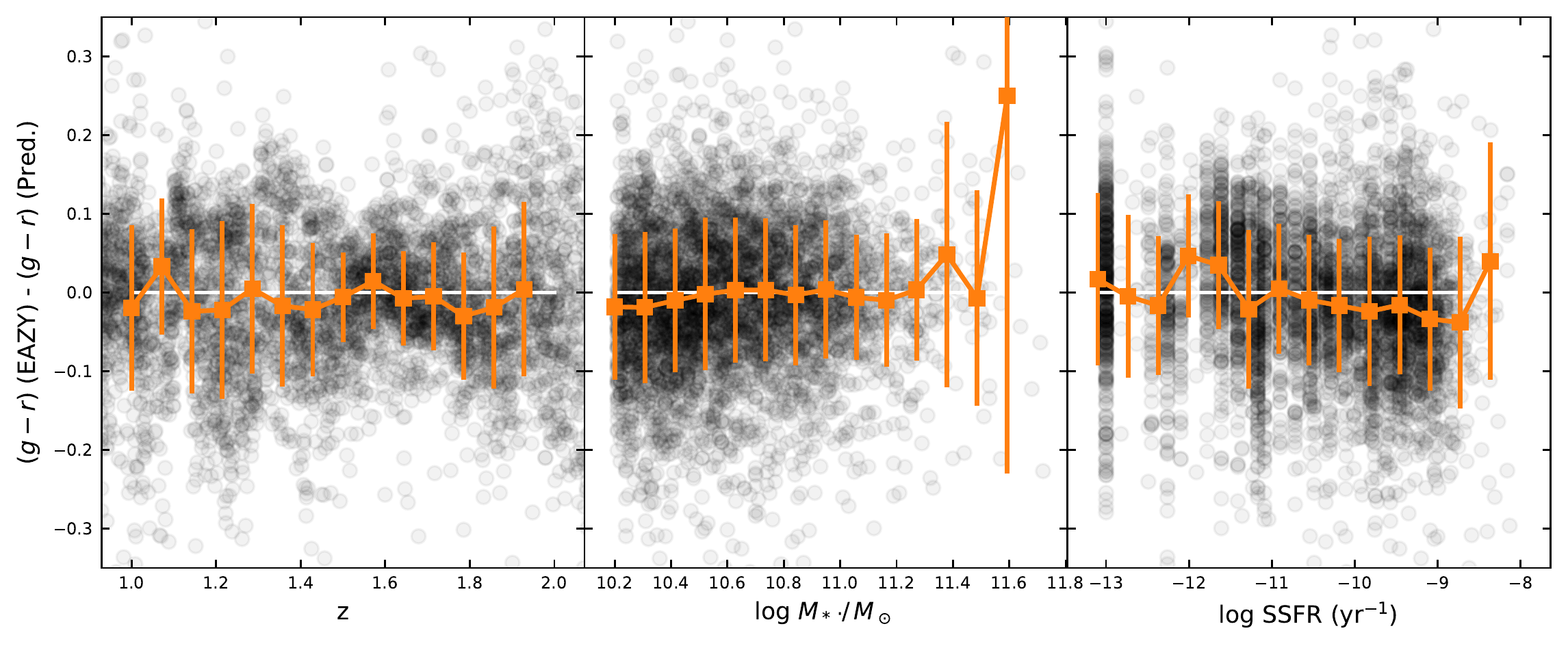}
    \includegraphics[width = 0.95\textwidth]{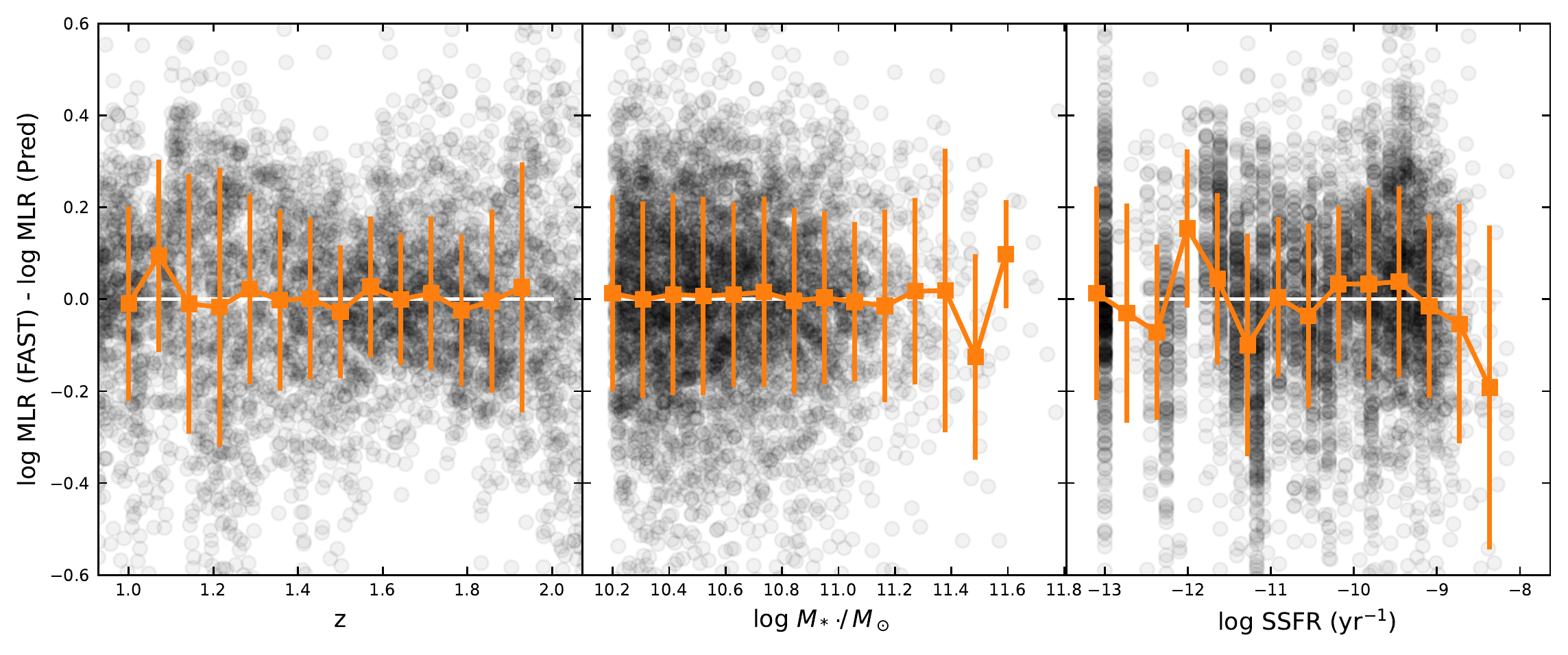}

    \caption{Residuals of the empirical fit performed to calculate rest frame $(g-r)$ (top) and \MLR{} (bottom) from the observed color as a function of galaxy properties. We see there are no large systematic effects as a for different galaxy populations for either fit. The orange lines show the median and scatter calculated with the bi-weight location and bi-weight scale.
    }
    \label{fig:col_fit_res}
\end{figure}

In Table~\ref{tab:obs_col_param} we show the best fit parameters for the empirical relations used in Section~\ref{sec:hm_meas} to convert from observed color to rest frame color or \MLR{}. The best fit parameters were found by maximizing the likelihood, taking into account measurement uncertainties in both the $x$ and $y$ parameters following the prescription in ~\citet{Hogg2010}. Separate fits were performed in six redshift bins with $dz = \pm 0.1$ and the central redshift listed in the table. We tested the alternative of fitting a smooth model, $ a = f(z)$ for e.g., instead of using separate redshift bins and found similar overall scatter. However when using linear or quadratic functions for the evolution we observed structure in the residuals as a function of $z$ so we opted to perform fits in separate bins instead. Also listed in the table is the RMS scatter for the best fit line.

Residuals to the empirical fit to both the rest from color and \MLR{} as a function of galaxy properties are shown in Figure~\ref{fig:col_fit_res}. The stellar masses and SSFR's are measured using FAST\citep{Skelton2014}. For clarity we show all galaxies with SSFR $< 10^{-13}\ {\rm yr}^{-1}$ at $10^{-13}\ {\rm yr}^{-1}$. We observe that there are no large systematic offset as a function of galaxy properties. 

\section{Comparison of half-mass radii to previous studies}
\label{sec:meth_comp}

\begin{figure}
    \centering
    \includegraphics[width = 0.95\textwidth]{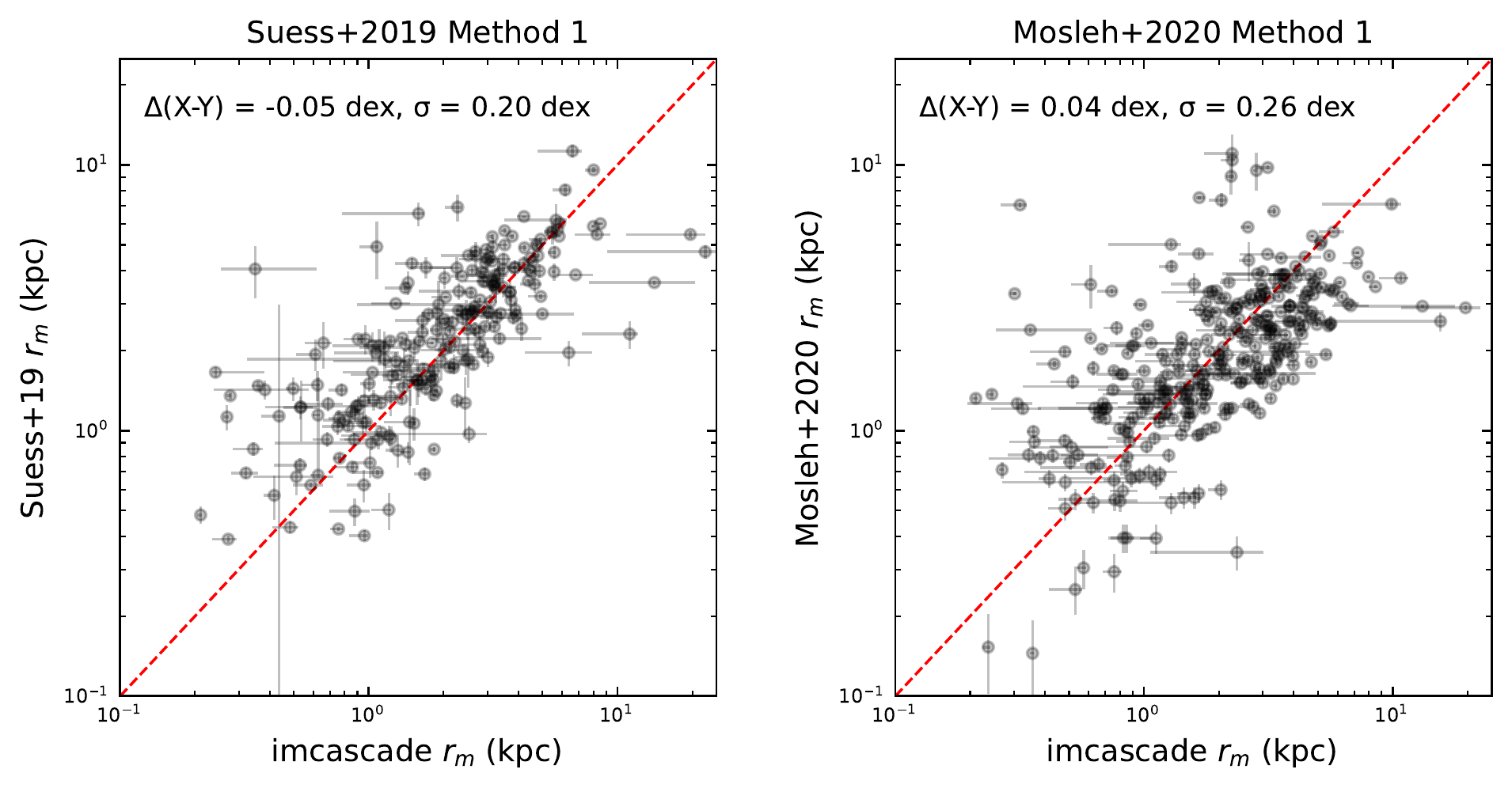}
    \includegraphics[width = 0.95\textwidth]{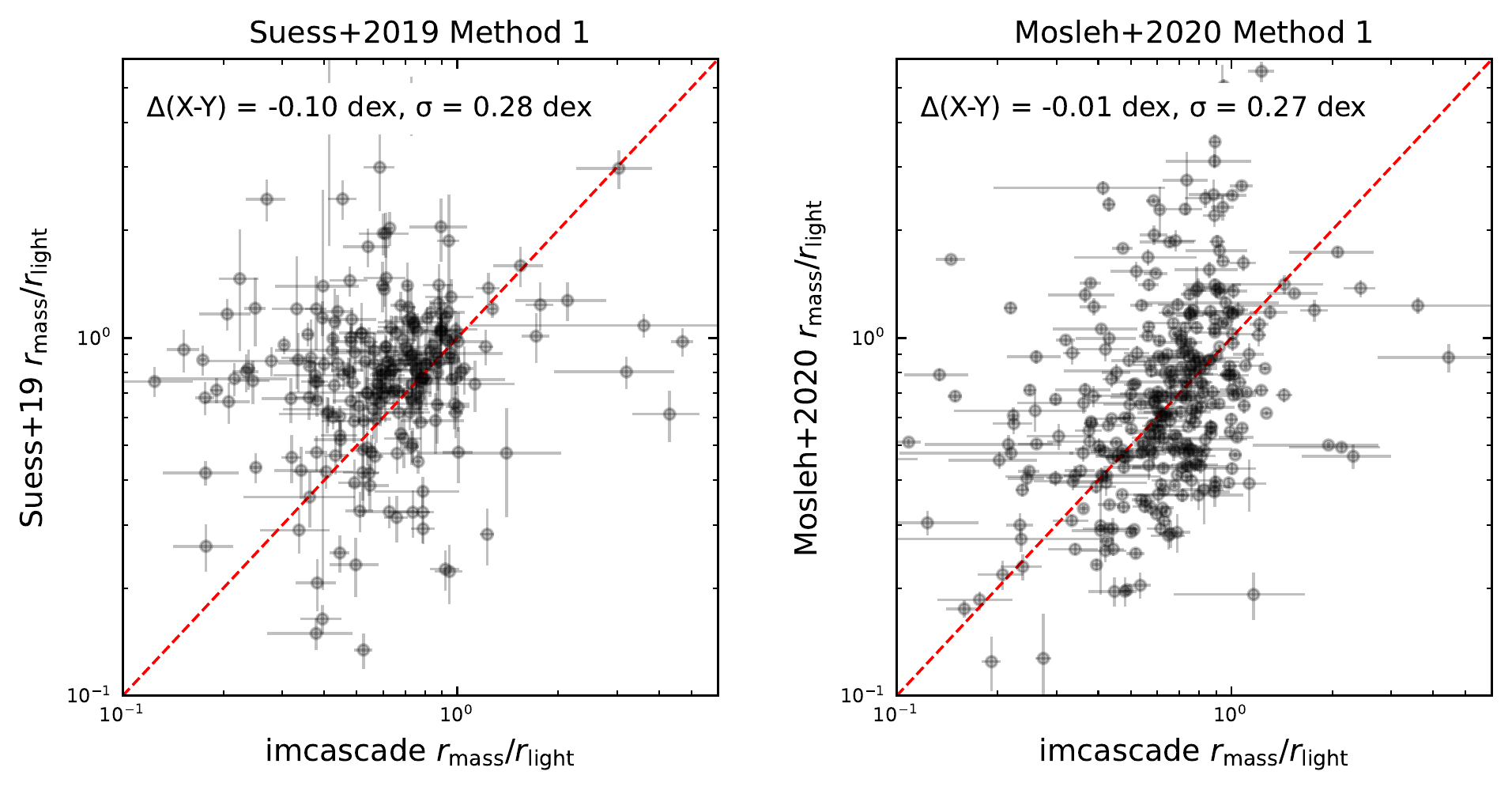}

    \caption{Comparing $r_{\rm mass}$ measurements from this study to \citet{Suess2019} and \citet{Mosleh2020}. The comparison of $r_{\rm mass}$ (top) and the ratio $r_{\rm mass}/r_{\rm light}$ (bottom) are shown. In each panel we show the average offset and scatter measured using the biweight location and biweight scale respectively. On average the three methods agree within $\sim 10\%$ yet there are systematic biases when comparing individual galaxies.
    }
    \label{fig:meth_comp}
\end{figure}

In this section we compare our measurements to those of ~\citet{Suess2019} and ~\citet{Mosleh2020}. Both studies test multiple methods, we opt to use Method 1 for each study as it is favored by each of the authors. In Method 1 of \citet{Suess2019} the authors start by performing SED fitting to calculate the convolved \MLR{} in concentric annuli and using ~\texttt{GALFIT} to find the best fit Sersic model to describe the light profile in the F160W image. They use a power-law parameterization for the \MLR{} profile which they multiply by the intrinsic light model and convolve with the PSF to construct a convolved mass map which they compare to the image of the galaxy and measure the convolved \MLR{} profile. This is then compared to the results of SED fitting to find the best fit \\MLR{}{}, which is multiplied by the Sersic light model to calculate the intrinsic mass map, where the half-mass radius is measured from. \citet{Mosleh2020} also perform SED fitting directly to pixels of the PSF matched CANDELS images. The result is the observed stellar mass map. This is used to measure a 1D mass density profile, which is fit with a 1D Sersic profile while accounting for the effect of the PSF.

Figure~\ref{fig:meth_comp} compares the half-mass radius measured using the different methods for overlapping galaxies in the GOODS-S field. This includes 247 galaxies in both our and the \citet{Suess2019} sample and 347 in the \citet{Mosleh2020} sample. On average our measurements agree with the previous studies to within $10\%$. This is reassuring given that the three different methods use vastly different methodologies to measure the half-mass radius. However there is significant scatter of $\sim 0.2-0.25$ dex between the methods. The median observed error for these galaxies is 0.1,0.1,0.15 for this study, \citet{Suess2019} and \citet{Mosleh2020}, respectively. This implies systematic uncertainties in these relations beyond the observational uncertainties. While the three methods agree on average, they do no appear to be entirely consistent on the level of an individual galaxy

Also shown is the comparison of $r_{\rm mass}/r_{\rm light}$. In some sense this is the more important comparison as the \MLR{} gradient is the key to measuring $r_{\rm mass}$ yet is difficult to measure and prone to systematic biases. We see very similar results when comparing $r_{\rm mass}$. While all 3 methods agree on average the observed scatter is larger than the observational uncertainties.

\section{Comparison of the \lowercase{$r_{\rm light}$} -stellar mass relation to previous studies}
\label{sec:size_mass_comp}
\begin{figure}
    \centering
    \includegraphics[width = 0.5\textwidth]{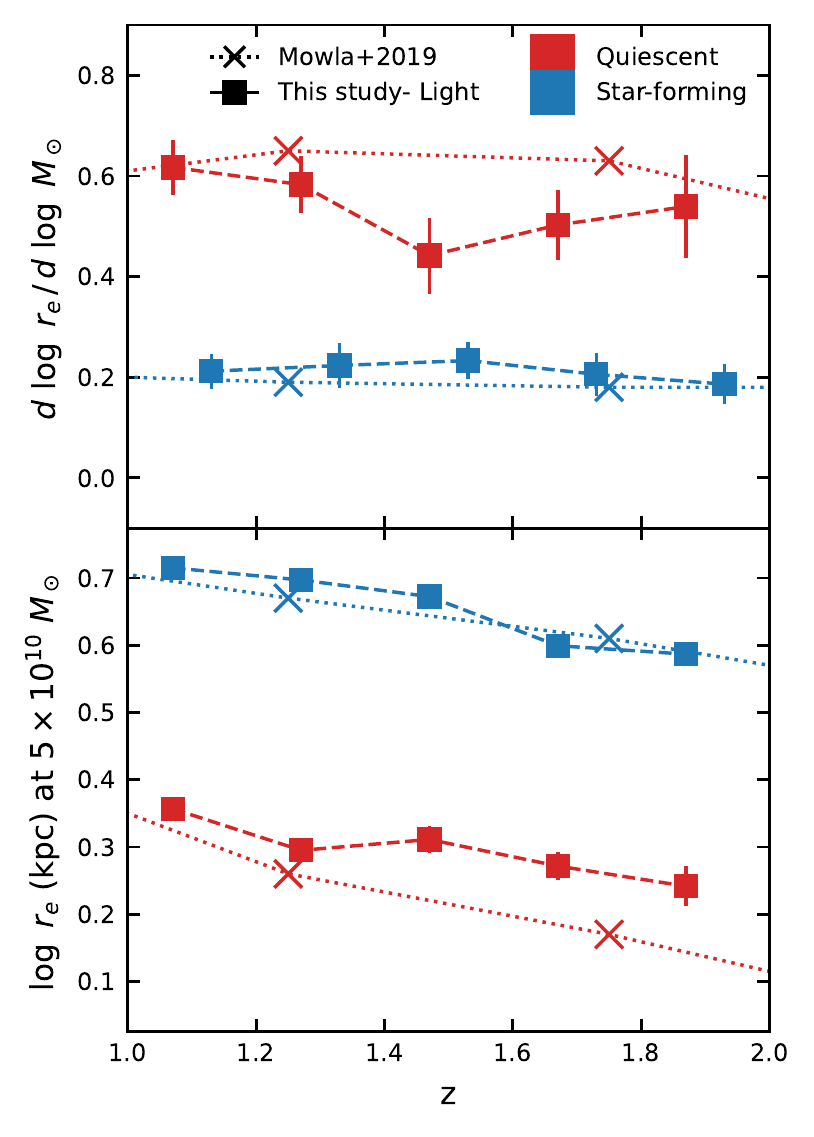}
    \caption{The best fit-parameters of the $r_{\rm light}- M_*$ relationship in this study compared to \citet{Mowla2019}. For star-forming galaxies our measurements agree well. For quiescent galaxies, the slope is generally similar but the normalization differs, especially at $z>1.5$. This is likely due to a difference in parameterization between \imc{}'s MGE models and the Sersic profiles us in \citet{Mowla2019} }
    \label{fig:M19_sm_comp}
\end{figure}

The best fit relationship to the $r_{\rm light}- M_*$ found in this study is compared \citet{Mowla2019} is shown in Figure~\ref{fig:M19_sm_comp}. The authors use the COSMOS-DASH survey, which covers a larger area compared to CANDELS, to measure the sizes of massive galaxies using~\texttt{GALFIT} assuming a single Sersic profile. This combined with data from ~\citet{vanderWel2014} to update the $r_{\rm light}- M_*$ relationship.

For star-forming galaxies, our two studies agree, within uncertainties, on the evolution of both the slope and normalization of the $r_{\rm light}- M_*$ relationship. For quiescent galaxies we see qualitatively consistent evolution however in detail there are a few discrepancies. Mainly at $z=1.5$ we measure a shallower slope, by 0.2, and at $z>1.5$ we find the half-light sizes of quiescent galaxies are $\sim 25\%$ larger on average then \citet{Mowla2019}. It is possible these difference are due to the different galaxies sample, our study only use the CANDELS survey and therefore contains far fewer galaxies at $\log M_*/ M_\odot > 11$. Another possibility is the difference between the MGE parameterization used in \imc~ and the Sersic profile. For quiescent galaxies the measured Sersic index is often greater than $4$ which produces profiles with very large central concentrations. This central peak is much greater than the MGE models which, for even the components with the smallest widths, become flat near the center. This would lead to the smaller sizes for the Sersic profiles. It is unclear which parameterization is more accurate for the central regions of these galaxies without higher resolution data; the half light radii are already on the order of the PSF FWHM.

\end{document}